\documentclass[aps,showpacs,amsmath,amssymb]{revtex4}
\usepackage{graphicx}
\usepackage{dcolumn}
\usepackage{bm}
\usepackage{epsfig}

\begin{document}
\title{Quasinormal modes of the Hayward black hole surrounded by quintessence: scalar, electromagnetic and gravitational perturbations}

\author{Omar Pedraza$^1$}
\email{omarp@uaeh.edu.mx}
\author{L. A. L\'opez$^1$}
\email{lalopez@uaeh.edu.mx}
\author{R. Arceo$^2$}
\email{roberto.arceo@unach.mx}
\author{I. Cabrera-Munguia$^{3}$}
\email{roberto.cabrera@uacj.mx, icabreramunguia@gmail.com}

\affiliation{$^1$ \'Area Acad\'emica de Matem\'aticas y F\'isica, UAEH, 
Carretera Pachuca-Tulancingo Km. 4.5, C. P. 42184, Mineral de la Reforma, Hidalgo, M\'exico.}

\affiliation{$^{2}$ Facultad de Ciencias en F\'isica y Matem\'aticas, Universidad Aut\'onoma de Chiapas, C. P. 29050, Tuxtla Guti\'errez, Chiapas, M\'exico.}

\affiliation{$^{3}$ Departamento de F{\'i}sica y Matem\'aticas, Universidad Aut\'onoma de Ciudad Ju{\'a}rez, 32310 Ciudad Ju\'arez, Chihuahua, M\'exico.}


\begin{abstract}
We study the quasi-normal modes for scalar, electromagnetic, and gravitational axial perturbations in the Hayward regular black hole surrounded by quintessence (HBH-$\omega_q$). Using the third--order WKB approximation we can determine the dependence of the quasi--normal modes on the parameters of the regular black hole and the parameters on the test fields. We also determine the greybody factor, giving transmission and reflection coefficients of the scattered wave through the effective potentials in the WKB approximation using numerical analysis.   
\\
\\
{\it Keywords:} Quasi--normal modes, Quintessence, WKB approximation.
\pacs{04.20.-q, 04.70.-s, 04.70.Bw, 04.20.Dw}
\end{abstract}

\maketitle
\section{Introduction}

When a black hole (BH) is perturbed, the resulting behavior can be described in three stages. The first stage corresponds to radiation due to the initial conditions of the perturbations. The second stage corresponds to damped oscillations with complex frequencies, the modes of such oscillations are called quasinormal modes (QNM). The third stage in general corresponds to a power law decay of the fields.

The frequencies of QNM of a BH are complex quantities that corresponds to solutions of the perturbed equations, which satisfy the boundary conditions of the purely ongoing wave at the horizon and the purely outgoing wave at infinity. In addition, its real part describes the real oscillation frequency and the imaginary part describes the damping of these oscillations. The importance of quasi-normal modes lies in the analysis of the stability of BHs, however, they also play another fundamental role in characterizing gravitational wave signals as the ones recently detected by LIGO  and VIRGO \cite{LIGOScientific:2016sjg}. 

Perturbations in BHs and their study have 
been taken into account for a long time  \cite{Regge:1957td,Zerilli:1970se,Teukolsky:1972my,Berti:2009kk,Konoplya:2011qq}. All the works on the study of QNM lead to a wave equation with a specific effective potential, depending on the characteristics of the effective potential in the literature propose several numerical methods to calculate QNM as the continued fraction method \cite{Percival:2020skc}, finite difference method \cite{Ma:2020qkd}, WKB  \cite{Schutz:1985km} approximation method, and the asymptotic iteration method (AIM) (see \cite{Cho:2011sf}). Furthermore, in the Eikonal limit is considered to study the relationship among unstable null geodesics,  Lyapunov exponents \cite{Cardoso:2008bp} and QNM.

Different investigations have emerged about QNM for a variety of scenarios. For example, in \cite{Devi:2020uac} the QNM for a Gauss--Bonnet de Sitter BH are studied, QNM of BH in general relativity coupled to nonlinear electrodynamics (NLED) have been studied in \cite{Toshmatov:2018ell, Toshmatov:2018tyo} where have they considered; scalar, electromagnetic and gravitational perturbations. Also, in \cite{Breton:2016mqh}, the behavior of QNM is shown to apply the Eikonal regime and effective geometry. The QNM of Hayward, Bardeen, and Ayón--Beato--García regular black holes are compared in \cite{Toshmatov:2015wga}.

Recent measurements show the accelerating expansion of the universe and that hypothetical "Dark Energy" dominates the universe, then the BHs surrounded by dark energy are of interest to researchers. There are alternative models as candidates for dark energy, most of them are based on a scalar field as the quintessence \cite{Capozziello_2006, PhysRevLett.81.3067}. For example, solutions to the Einstein equations with the assumption of spherical symmetry with quintessence were obtained by Kiselev \cite{Kiselev:2002dx}, and there have been different investigations \cite{Ghaderi:2017wvl, Fernando:2012ue, Malakolkalami:2015tsa, Ghosh:2016ddh} of BHs surrounded by the quintessence applying the Kiselev model. Moreover, the QNM of the regular Bardeen BH surrounded by quintessence are studied in \cite{Saleh:2018hba}.

The regular solutions have been proposed in NLED, another idea to propose regular solutions is to consider that a regular solution will contain critical scale, mass, and charge parameters restricted by some value, which depends only on the type of the curvature, this assumption, is called the limiting curvature conjecture \cite{Polchinski:1989ae}.
Following the idea of the limiting curvature Hayward \cite{Hayward:2005gi} proposed a static spherically symmetric BH that near the origin behaves like a de Sitter space--time, its curvature invariants being everywhere finite and satisfying the weak energy condition. Rotating Hayward has been studied in \cite{Amir:2015pja}, and Hayward charged in \cite{Frolov:2016pav}. Also, \cite{Lopez, Lin:2013ofa} studied the QNM  of Hayward BH. The Geodesics structure for Hayward BH surrounded by quintessence in \cite{Pedraza:2020uuy}.

Motivated by the above mentioned, in the present paper, we studied the effects of quintessence at the behavior of  QNM for the scalar, electromagnetic and gravitational perturbations of Hayward BH. The paper is organized as follows. In Sec. II, the Hayward BH surrounded by quintessence is presented, and we briefly discuss the event horizons. Sec III, we describe the scalar, electromagnetic and gravitational perturbations of a BH. We analyze the behavior of effective potential for different perturbations considering the special cases when the quintessence state parameter takes the values $-2/3$ and $-4/9$. Then, the QNMs are introduced in Sec. IV and we analyze them by using a third--order WKB approximation. The reflection and transmission coefficients are studied considering the different perturbation in Sec. V. Conclusions are given in the last section.

\section{A Hayward black hole surrounded by quintessence}\label{sec.hori}

Kiselev \cite{Kiselev:2002dx}  proposed static and spherically symmetric solutions that describe BHs surrounded by quintessence. We focus our attention to the static, spherically symmetric Hayward BH \cite{Hayward:2005gi} surrounded by quintessence, described by the line element \cite{Pedraza:2020uuy}

\begin{equation}\label{mfa}
ds^2=-f_{\omega_q}(r)dt^2+\frac{dr^2}{f_{\omega_q}(r)}+r^2d\theta^2+r^2\sin^2\theta d\phi^2\,,
\end{equation}
where
\begin{equation}\label{ec.rfc}
f_{\omega_q}(r)=1-\frac{2Mr^2}{r^3+2M\epsilon^2}-\frac{c}{r^{3\omega_q+1}}\,.
\end{equation}

Here, $M$ is the mass and $\epsilon$ is a parameter related to the cosmological constant, $c$ is a normalization factor and $\omega_q$ is the quintessence state parameter which has the range $-1<\omega_q<-1/3$.

In our preceding paper \cite{Pedraza:2020uuy}, we analyzed in detail the null geodesics for different energies as well as the behavior of horizons. To complement the study of horizons, we can study the positive roots of $f_{\omega_q}(r)=0$. This condition leads to the polynomial

\begin{equation}\label{ec.dnh}
r^{3\omega_q+4}-2r^{3\omega_q+3}+2\varepsilon^2r^{3\omega_q+1}-cr^3-2c\varepsilon^2=0\,,
\end{equation}   

where, we have expressed the term $\epsilon$, the quintessential parameter and radial distance in units of the $M$, $\varepsilon\to \varepsilon/M$, $c\to c/M^{3\omega_q+1}$, $r\to r/M$. Of course, the number of horizons depends on the values of parameters $\omega_q$, $\varepsilon$ and $c$, i. e. we could have a BH, an extremal BH, or a naked singularity. Following the method employed in Ref. \cite{Rizwan:2018lht}, by means of (\ref{ec.dnh}) we can parametrize $\varepsilon^2$ as a function of $r$ and $c$ as;

\begin{equation}\label{pec.ep}
\varepsilon^2(r,c)=-\frac{r^2}{2}
\frac{r^{3\omega_q+2}-cr-2r^{3\omega_q+1}}{r^{3\omega_q+1}-c}\,,
\end{equation} 

where $\varepsilon^2$ has extrema $\left (\frac{d \varepsilon^2 }{dr}\arrowvert_{c_e}=0\right )$ for $c=c_e$, given by;

\begin{equation}\label{ec.cex}
c_e(r)=c_{\pm}=r^{3\omega_q}\left[r-\omega_q-1
\pm\frac{1}{\sqrt{3}}\sqrt{
3\omega_q^2+6\omega_q+3-2r(3\omega_q+1)
}\right]
\,.
\end{equation}

The critical value of $c_{\pm}$ is located at;

\begin{equation}
r_{c}=
\frac{9\omega_q^2+42\omega_q+8+\left(3\omega_q-2\right)\sqrt{9\omega_q^2+16}}
{12\left(3\omega_q+1\right)}
\,.
\end{equation}
So, the critical values of $\varepsilon^2_{crit}$ and $c_{crit}$ are given by
\begin{eqnarray}
\varepsilon^2_{crit}=\varepsilon^2_{\pm}&=&-\frac{r_c^2}{2}
\frac{r_c^{3\omega_q+2}-c_{\pm}r_c-2r_c^{3\omega_q+1}}{r_c^{3\omega_q+1}-c_{\pm}}\,,\label{epcriw}
\\
c_{crit}=c_{\pm}&=&r_c^{3\omega_q}\left[r_c-\omega_q-1
\pm\frac{1}{\sqrt{3}}\sqrt{
3\omega_q^2+6\omega_q+3-2r_c(3\omega_q+1)
}\right]
\,.\label{ccriw}
\end{eqnarray}

\begin{figure}[!h]
\centering
\includegraphics[scale=0.3]{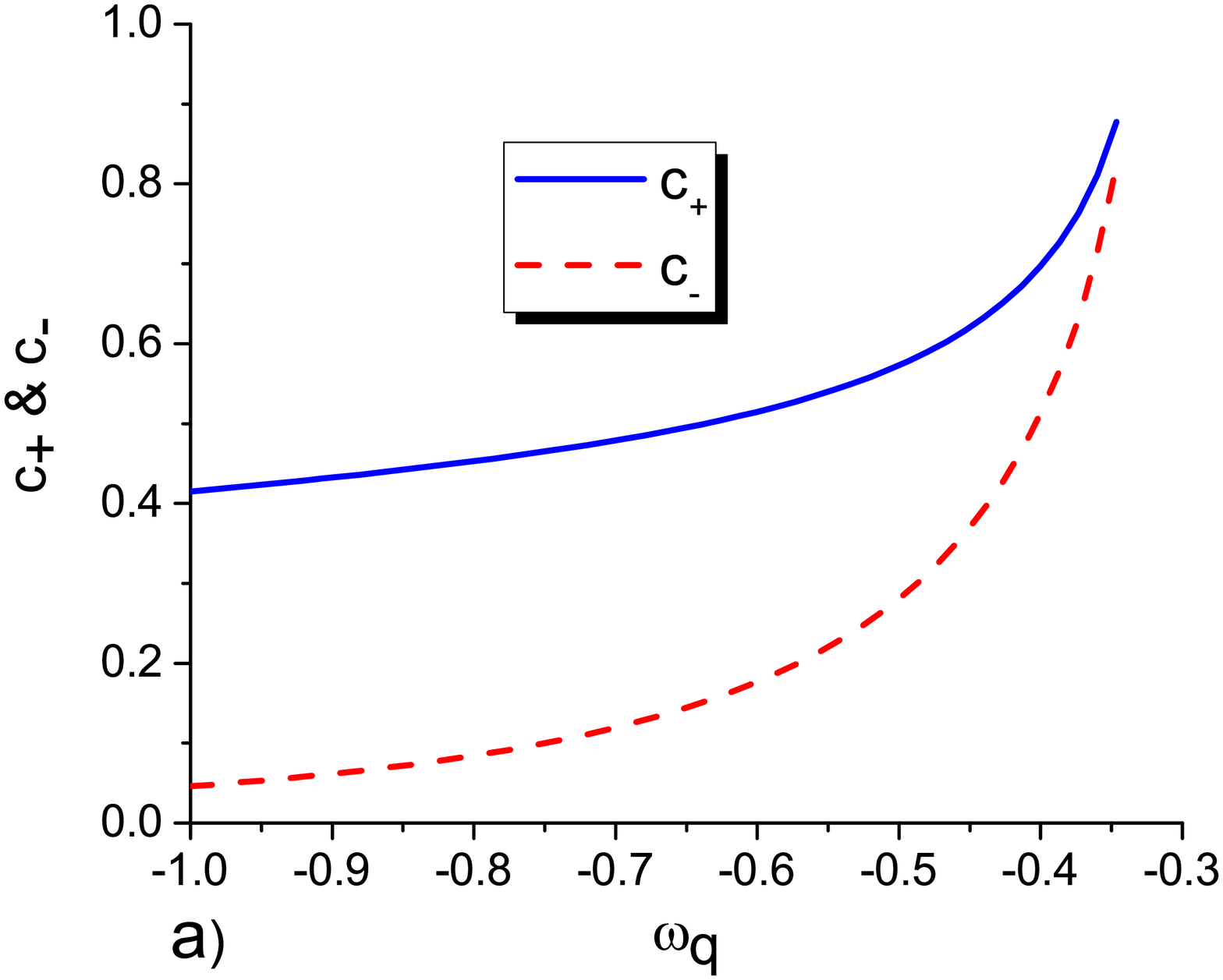}
\includegraphics[scale=0.3]{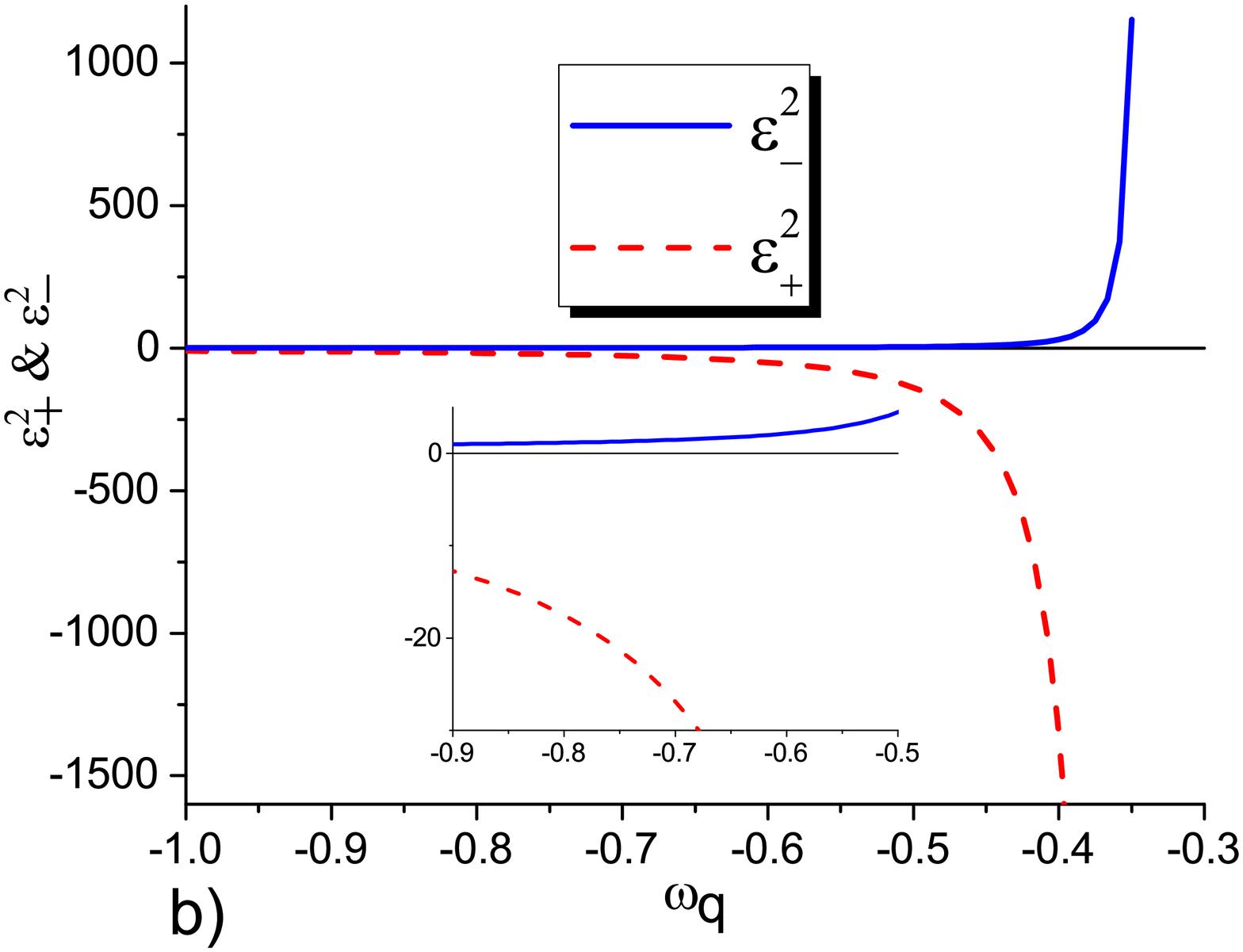}
\caption{$a)$ In this figure, we show the critical values of the quintessential parameter $c_+$ and $c_-$. $b)$ In this figure, we plot $\varepsilon^2_+$ and $\varepsilon^2_-$ as a function of $ \omega_q $.}
\label{fmccc}
\end{figure}

In Fig. \ref{fmccc}, we can see that $c_+$ and $c_-$ are positive quantities and both of them increase if $\omega_q$ increases as well, while $\varepsilon^2_+$ is a positive quantity and for $\omega_q\to-1$, we get $\varepsilon^2_+\to 0.4$, when $\omega_q\to-1/3$, we obtained that $\varepsilon^2_+\to\infty$. On the other hand, $\varepsilon^2_-$ is a negative quantity and lacks physical sense.

In summary for  $0\leq \varepsilon^2 \leq \varepsilon^2_{-}$ and $0 \leq c \leq c_{-} $ the Hayward BH surrounded by quintessence (HBH--$\omega_q$) can represent a BH with different horizons. The number of horizons depend entirely on the choice of the values of parameters, different authors (for example \cite{Fernando:2012ue} \cite{Malakolkalami:2015tsa}) have observed that when considering the quintessence term, a new horizon emerges, the cosmological (quintessence) horizon. Then as Hayward BH has two horizons, Hayward BH--$\omega_q$ could have three horizons; $r_{in}$, $r_{out}$ and $r_{\omega_{q}}$ (quintessence horizon). 

In Table \ref{Tab1} we display some values of $c_c$ and $\varepsilon^2_c$ for different values of $\omega_q$, it is clear how the factor $\omega$ modifies the behavior of the horizons.

\begin{table}[!hbt]
	\begin{center}
		\begin{tabular}{|c|c|c|c|c|c|c|}
			\hline
			$\omega_q$ & $-4/9$ &  $-1/2$ & $-5/9$ & $-2/3$ & $-7/9$ & $-8/9$\\
			\hline
			$c_{c}$ &0.381984 &  0.280663 & 0.214634 & 0.135255 & 0.090882 & 0.063723\\
			\hline
			$\varepsilon^2_{c}$ & 9.684011&  4.480419  & 2.809328 & 1.642988& 1.224015 & 1.020501\\
			\hline
		\end{tabular}
		\caption{Critical values of the $c$ and $\varepsilon^2$ a for different $\omega_q$}\label{Tab1}
	\end{center}
\end{table}

For any value of $\omega_q$ the HBH--$\omega_q$ extremal (when two or more horizons collapse into one) can be obtained from the conditions 

\begin{equation}\label{ec.df}
	\frac{d}{dr}f_{\omega_q}(r)=\frac{
		2r^{3(\omega_q+2)}-8\varepsilon^2r^{3(\omega_q+1)}+(3\omega_q+1)cr^6+4(3\omega_q+1)c\varepsilon^2r^3+4(3\omega_q+1)c\varepsilon^4}
	{r^{3\omega_q+2}\left(r^3+2\varepsilon^2\right)^2}
	=0\,,
\end{equation}  

where $f_{\omega_q}(r)=0$ is satisfied simultaneously, then introducing $\varepsilon^2$ of (\ref{pec.ep}) in (\ref{ec.df}), we get the condition
\begin{equation}\label{ec.rec}
	3r^{6\omega_q+3}-4r^{6\omega_q+2}-6cr^{3\omega_q+2}+6c(\omega_q+1)r^{3\omega_q+1}+3c^2r=0\,.
\end{equation}

With the aim to solve (\ref{ec.rec}), next we are going to consider some specific values for $\omega_q$.

\subsection{HBH--$\omega_q$ with  $\omega_q=-2/3$ and  $\omega_q=-4/9$}

\begin{figure}[!h]
\centering
\includegraphics[scale=0.28]{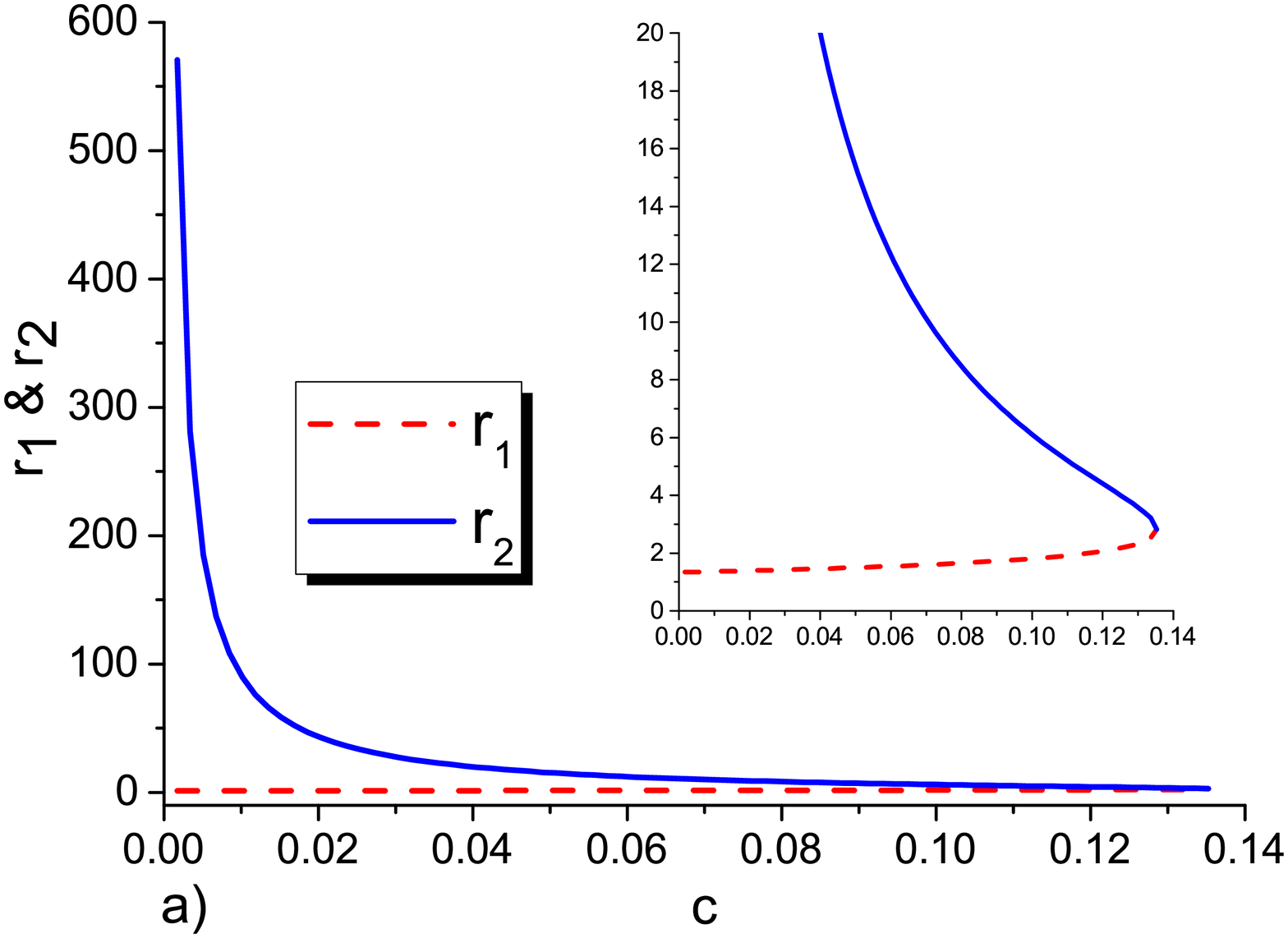}
\includegraphics[scale=0.28]{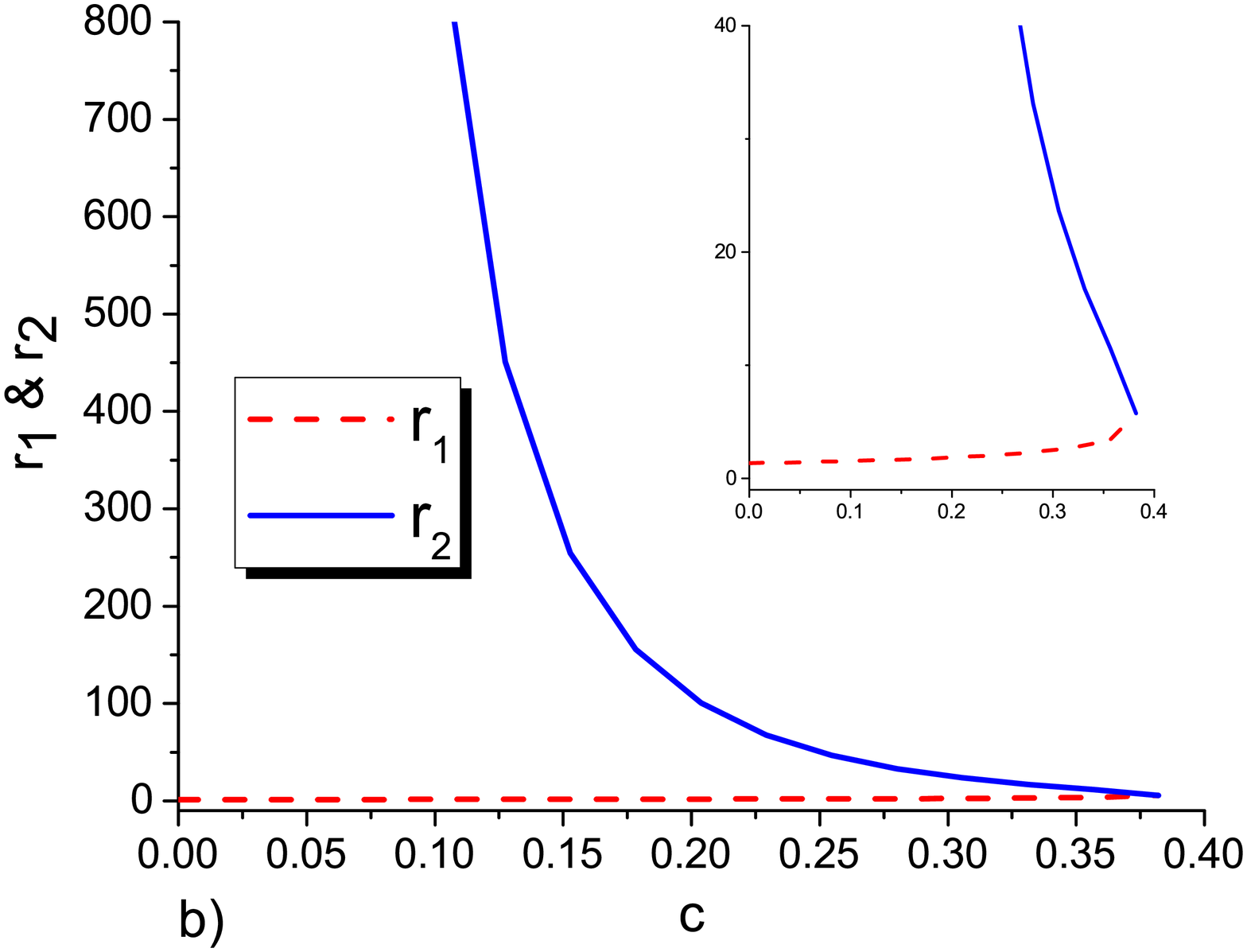}
\caption{The figure shows $r_1$ and $r_2$ as function of $c$.  a) $\omega_q=-2/3$ and b) $\omega_q=-4/9$ .}
\label{r1r2w23}
\end{figure}

\begin{figure}[!h]
\centering
\includegraphics[scale=0.28]{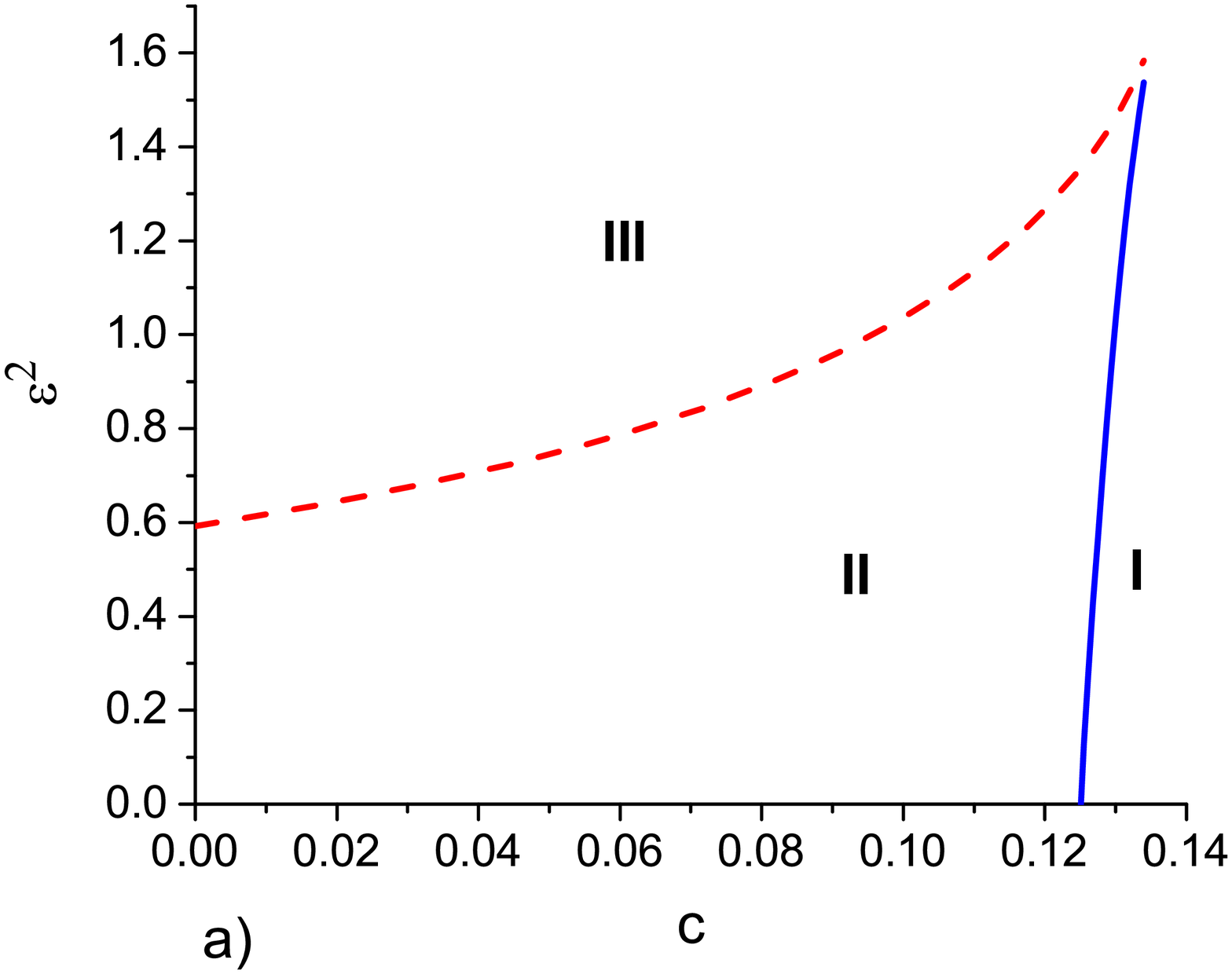}
\includegraphics[scale=0.28]{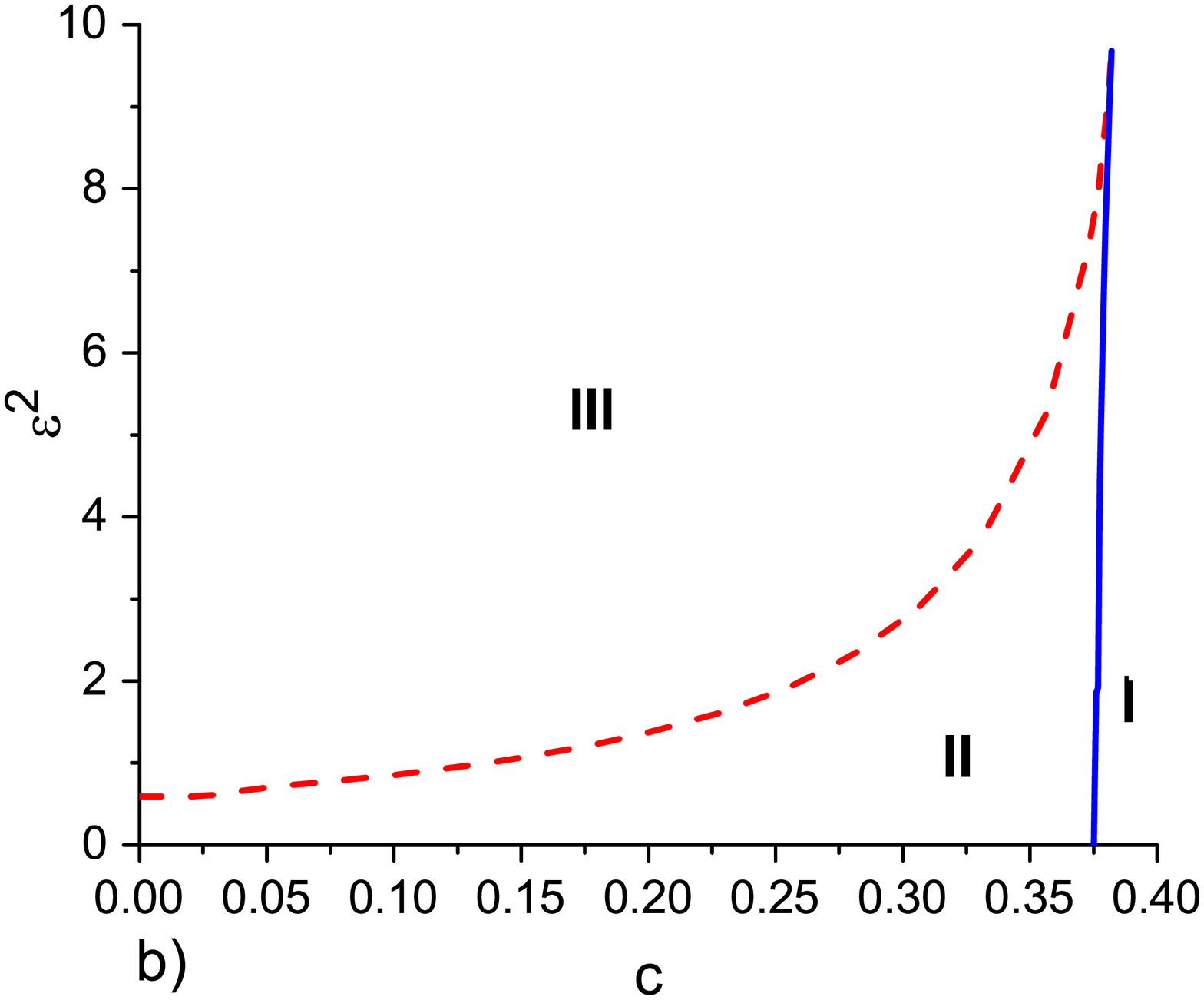}
\caption{The graphs show $\varepsilon^2$ as function of $c$. Region II represents a BH with three horizons. The boundary of regions II and III represents extremal BH of type I. The boundary of regions II and I represents extremal BH type II. For $c_c$ and $\varepsilon_c$, we have a BH type III. For any $(c,\varepsilon)$ in regions I and III, the BH presents naked singularities. a) $\omega_q=-2/3$ and b) $\omega_q=-4/9$}
\label{ec23}
\end{figure}

\begin{figure}[!h]
\centering
\includegraphics[scale=0.3]{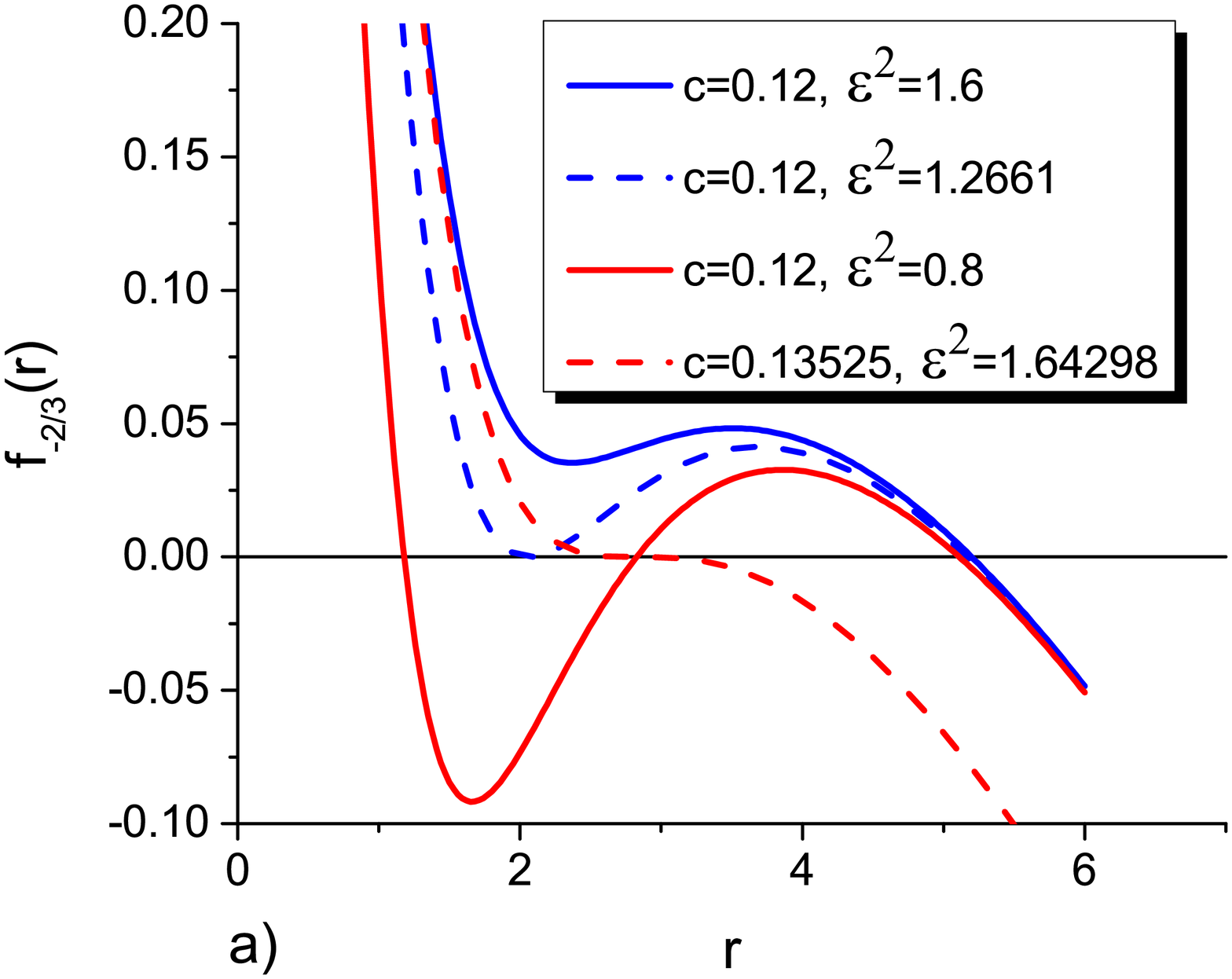}
\includegraphics[scale=0.3]{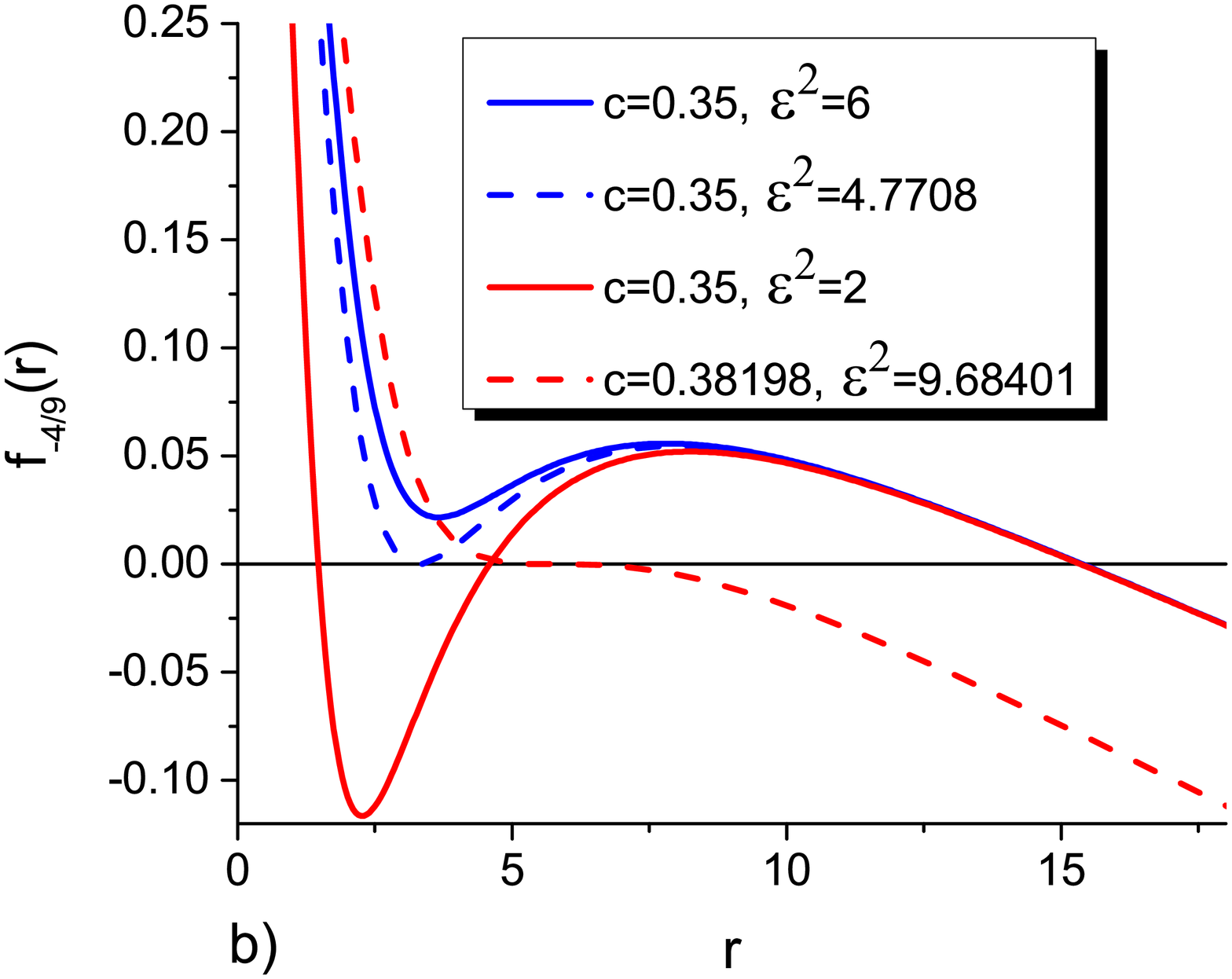}
\caption{The graph a) shows the relation of $f_{-2/3}(r)$ with $r$ for various values of $(\varepsilon^2,c)$ and b) shows the relation of $f_{-4/9}(r)$ with $r$ for various values of $(\varepsilon^2,c)$.}
\label{fm23}
\end{figure}

In this subsection, we study the behavior of extremal BHs and the naked singularities, we consider the particular cases $\omega_q=-2/3$ and $\omega_q=-4/9$. Both cases enable a relatively simple treatment of the properties of HBH-$\omega_q$.
For $\omega_q=-2/3$, Eq. (\ref{ec.rec}) becomes;

\begin{equation}\label{ec.ecw23}
3c^2r^3-6cr^2+2cr+3r-4=0\,.
\end{equation}
This expression can be solved analytically and has two real positive roots for $c\leq c_c$. The real positive roots are assume the form
\begin{eqnarray}
r_1&=&
\frac{2}{3c}+\frac{1}{6c}\left[
\frac{2c-1}{\Delta}-\Delta
\right]
+\frac{i}{2c\sqrt{3}}\left[
\Delta+\frac{2c-1}{\Delta}
\right]
\,,\label{ec.r1w23}\\
r_2&=&\frac{2}{3c}+\frac{1}{6c}\left[
\frac{2c-1}{\Delta}-\Delta
\right]
-\frac{i}{2c\sqrt{3}}\left[
\Delta+\frac{2c-1}{\Delta}
\right]
\,,\label{ec.r2w23}
\end{eqnarray}
with
\begin{equation}
\Delta=
\left[
\sqrt{8c^3+132c^2-18c}+12c-1
\right]^{1/3}
\,.
\end{equation}
From Fig. \ref{r1r2w23} a), we can see that as $c$ increases $r_1$ decreases and $r_2$ increases, and in Fig. \ref{ec23} a), the behavior of $\varepsilon^2$ as function of $c$ is shown. In region II, we have a BH--$\omega_q$ with three horizons. The boundary between I and II  and the boundary between II and III represents extremal BH--$\omega_q$, for the region I and III the HBH--$\omega_q$ have one horizon. On the other hand, Fig. \ref{fm23} a), shows the metric function for $\omega_q=-2/3$. Depending on the choice of the values of the parameters $\varepsilon^2$ and $c$ we can have a BH with one, two or three horizons.


For all values of ($c$, $\varepsilon^2$) in region II of Fig. \ref{ec23} a), there are three real positive roots of Eq. (\ref{ec.dnh}), $r_{in}$, $r_{out}$ and $r_q$ as shown in Fig. \ref{fm23} a). These horizons and extrema given by (\ref{ec.ecw23}) satisfy the relation $r_{in}\leq r_1\leq r_{out}\leq r_2\leq r_q$, on the other hand, from Fig. \ref{r1r2w23} a), we can see that for small values of $c$, $r_2$ become very large, and hence $r_q$ is very large too. So we can expect three types of extremal black holes.
\begin{enumerate}
\item
Type I: In this type of extremal BH--$\omega_q$, $r_1$ is the horizon of the black hole and is given by Eq. (\ref{ec.r1w23}), therefore $r_{in}=r_{out}$. This case is shown by the dashed line in Fig. \ref{ec23} a). Here $c\in(0,0.135255)$.  
\item 
Type II: Here, the second type of extremal BH--$\omega_q$, $r_2$ is the horizon of the black hole and is given by (\ref{ec.r2w23}), where $r_{out}=r_q$. This case corresponds to the continuous line in Fig. \ref{ec23} a). Note that $c$ is defined only for $0.125\leq c\leq 0.135255$.  
\item
Type III: I the third type of extremal BH--$\omega_q$, all three horizons merge into a single horizon, thus, we have $r_1=r_2$. In the which is possible for $c=c_c=0.135255$ and $\varepsilon^2=\varepsilon_c^2=1.642988$, as shown in Table \ref{Tab1}.

.   
\end{enumerate}  
 
For $\omega_q=-4/9$, Eq. (\ref{ec.rec}) can be written as 
\begin{equation}\label{dc.fm12}
3c^2r^{11/2}+3r^{9/2}-4r^{7/2}-6cr^5+3cr^4=0
\,.
\end{equation}
Unfortunately, Eq. (\ref{dc.fm12}) cannot be solved analytically, so we proceed numerically to solve this equation. By numerical analysis, we see that Eq. (\ref{dc.fm12}) has two real positive roots for $c\leq c_c=0.381984$, as shown in Fig.\ref{r1r2w23} b). 

Again, from Fig. \ref{ec23} b), the behavior of $\varepsilon^2$ with $c$ is shown. In this graph, we can see from the continuous line, that value of $c$ varies in a small interval. In the dashed line, the value of $\varepsilon^2$ is defined only for $0\leq c\leq 0.381984$. From Fig. \ref{fm23} b), we plot the metric function $f_{-4/9}(r)$ against $r$ for different values of $(\varepsilon^2,c)$. Comparing the different regions, it is possible to mention that $II_{\omega=-2/3} < II_{\omega=-4/9}$ and the range of $c$ for $\omega=-2/3$ is less than the range for $\omega=-4/9$. From the Fig. \ref{fm23} we can conclude that the quintessence horizon is very large.

\section{Perturbation equations in a HBH--$\omega_q$}

The general perturbation equation for the massless scalar field in the curved space--time is given by Klein--Gordon equation
\begin{equation}\label{ec.sfkg}
\frac{1}{\sqrt{-g}}\partial_{\mu}\left(\sqrt{-g}r^{\mu\nu}\partial_{\nu}\right)\Phi=0\,.
\end{equation}
Substituting Eq. (\ref{mfa}) into Eq. (\ref{ec.sfkg}), and using the ansatz for the scalar field $\Phi$  
\begin{equation}
\Phi=e^{-i\omega t}Y(\theta,\phi)\frac{\xi(r)}{r}\,.
\end{equation}
After introducing the tortoise coordinates change
\begin{equation}
dr_*=\frac{dr}{f_{\omega_q}(r)}\,,
\end{equation}
we obtain the radial perturbation equation 
\begin{equation}\label{ecmcnsf}
\frac{d^2\xi(r)}{dr_*^2}+\left[\omega^2-V_{\omega_q}(r)\right]\xi(r)=0\,,
\end{equation}  
where
\begin{equation}\label{ec.vscalar}
V_{\omega_q}(r)=f_{\omega_q}(r)\left[\frac{l(l+1)}{r^2}+\frac{f'_{\omega_q}(r)}{r}\right]\,.
\end{equation}
Here, $l$ is the spherical harmonic index. We can see from Eq. (\ref{ec.vscalar}) that the effective potential $V_{\omega_q}$ is dependent on the parameter $\varepsilon^2$, the quintessential term $c$, the quintessence state parameter $\omega_q$, and the angular harmonic index $l$. The generalized form of the effective potential for electromagnetic and gravitational test fields, can be written as
\begin{equation}\label{ec.poseg}
V_{\omega_q}(r)=f_{\omega_q}(r)\left[\frac{l(l+1)}{r^2}+\left(1-s^2\right)\frac{2m(r)}{r^3}
+\left(1-s\right)\left(\frac{f'_{\omega_q}(r)}{r}-\frac{2m(r)}{r^3}
\right)
\right],
\quad 
m(r)=\frac{Mr^3}{r^3+2\epsilon^2}+\frac{c}{2r^{3\omega_q}}\,,
\end{equation}
where $l$ is restricted by $l\geq s$, and $s=0,1,2$ denotes the spin of the perturbation: scalar, electromagnetic, and gravitational (see \cite{Medved:2004tp,PhysRevD.71.124033}). The effective potential $V_{\omega_q}(r)$ has asymptotic value $V_{\omega_q}(r) \approx r^{-6 \omega_q -4} \left(c^2 s^2+3 c^2 s \omega_q -3 c^2 \omega_q -c^2\right)$ when $r\to\infty$, in the case of electromagnetic perturbation $V_{\omega_q}(r) \approx 0$, for scalar and gravitational perturbation $V_{\omega_q}(r) \approx r^{-6 \omega_q -4} c^2$.

\begin{figure}[!h]
\centering
\includegraphics[scale=0.3]{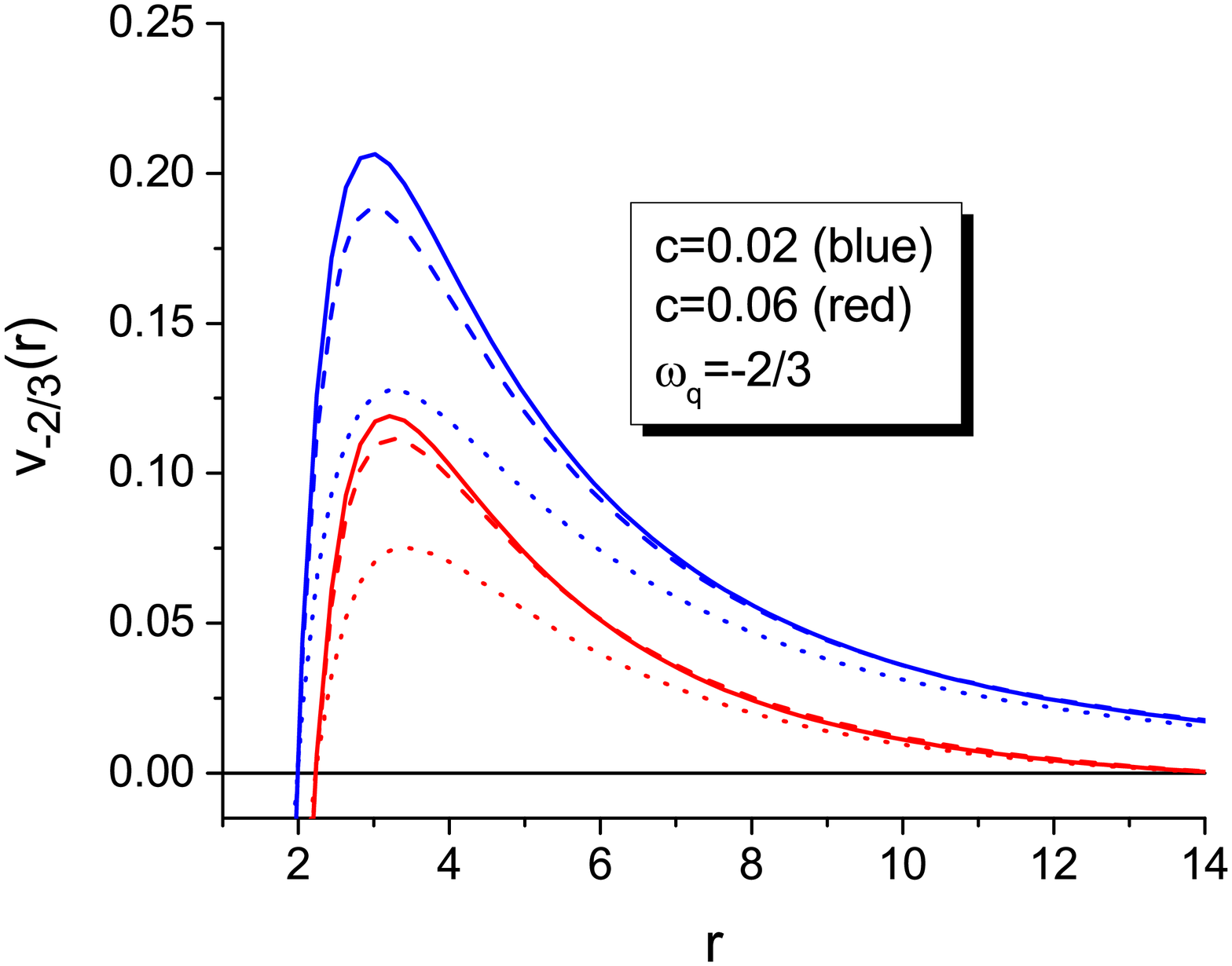}
\includegraphics[scale=0.3]{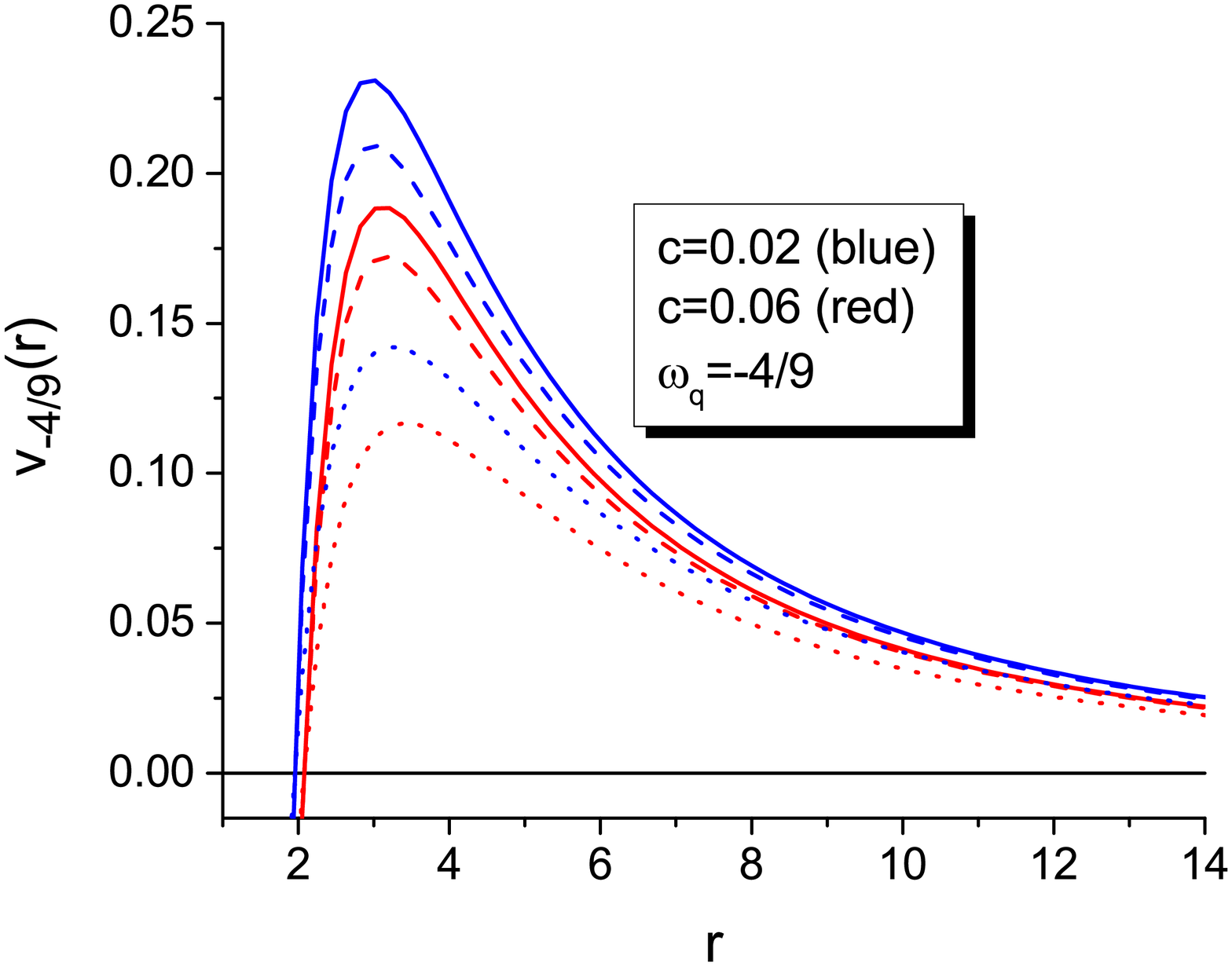}
\caption{The behavior of the effective potential of the scalar (solid), electromagnetic (dashed), and gravitational (dotted) perturbative fields with $c$, for $\varepsilon^2=0.2$, $l=2$ and $n=0$.}
\label{vscw}
\end{figure}

We plot the effective potential of the three kinds of perturbations. Fig \ref{vscw} shows the dependence of effective potential (\ref{ec.poseg}) on the parameter $c$ for $\omega_q=-2/3$ and $\omega_q=-4/9$, respectively. We can see that the potentials decrease with increasing $c$ and its position moves along the right then implies that the presence of quintessence reduces the magnitudes of the different potentials (\ref{ec.poseg}). Also, we can mention that in both cases $V_{\omega_q}(r)_{grav}< V_{\omega_q}(r)_{elec}< V_{\omega_q}(r)_{sc}$.

\begin{figure}[!h]
\centering
\includegraphics[scale=0.3]{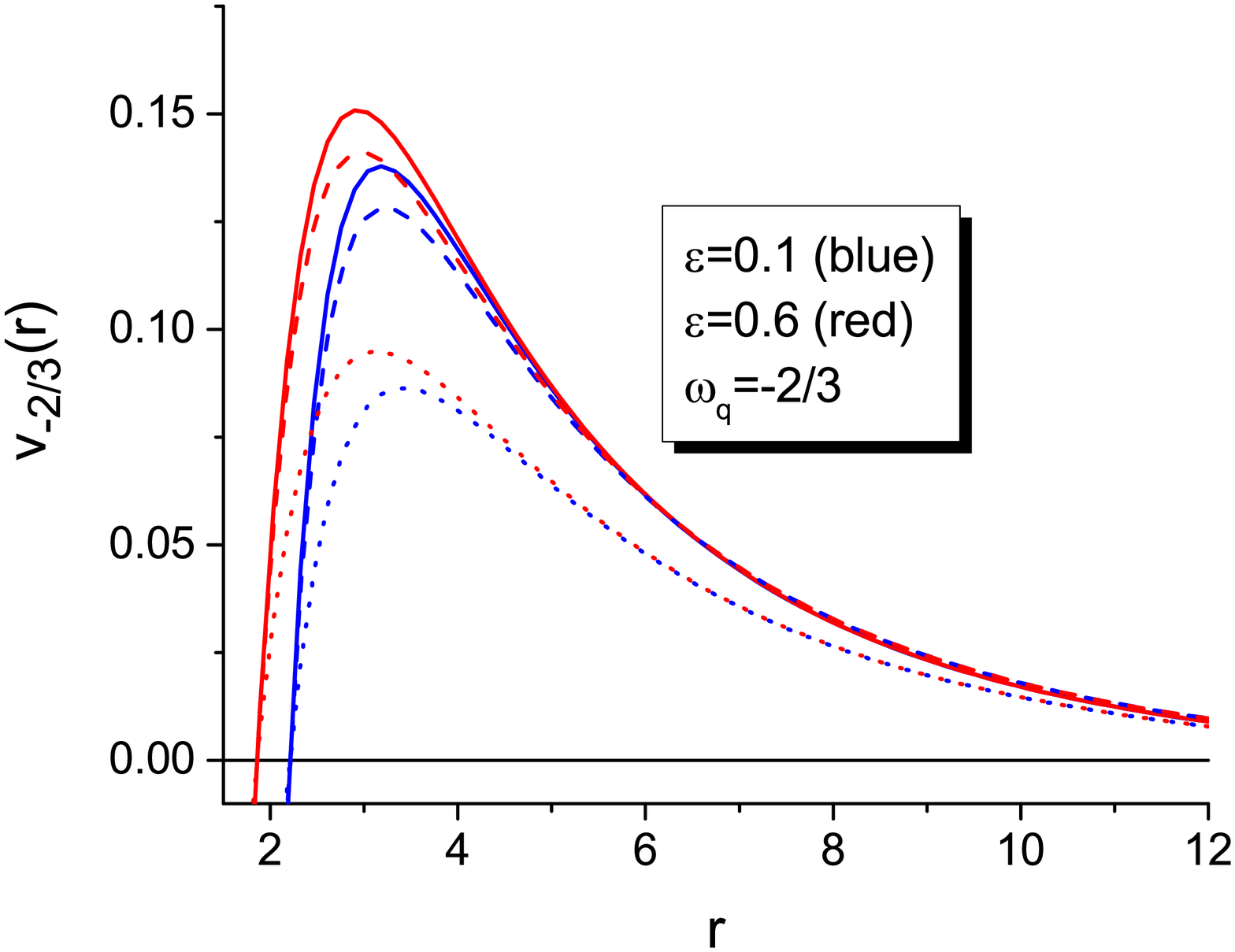}
\includegraphics[scale=0.3]{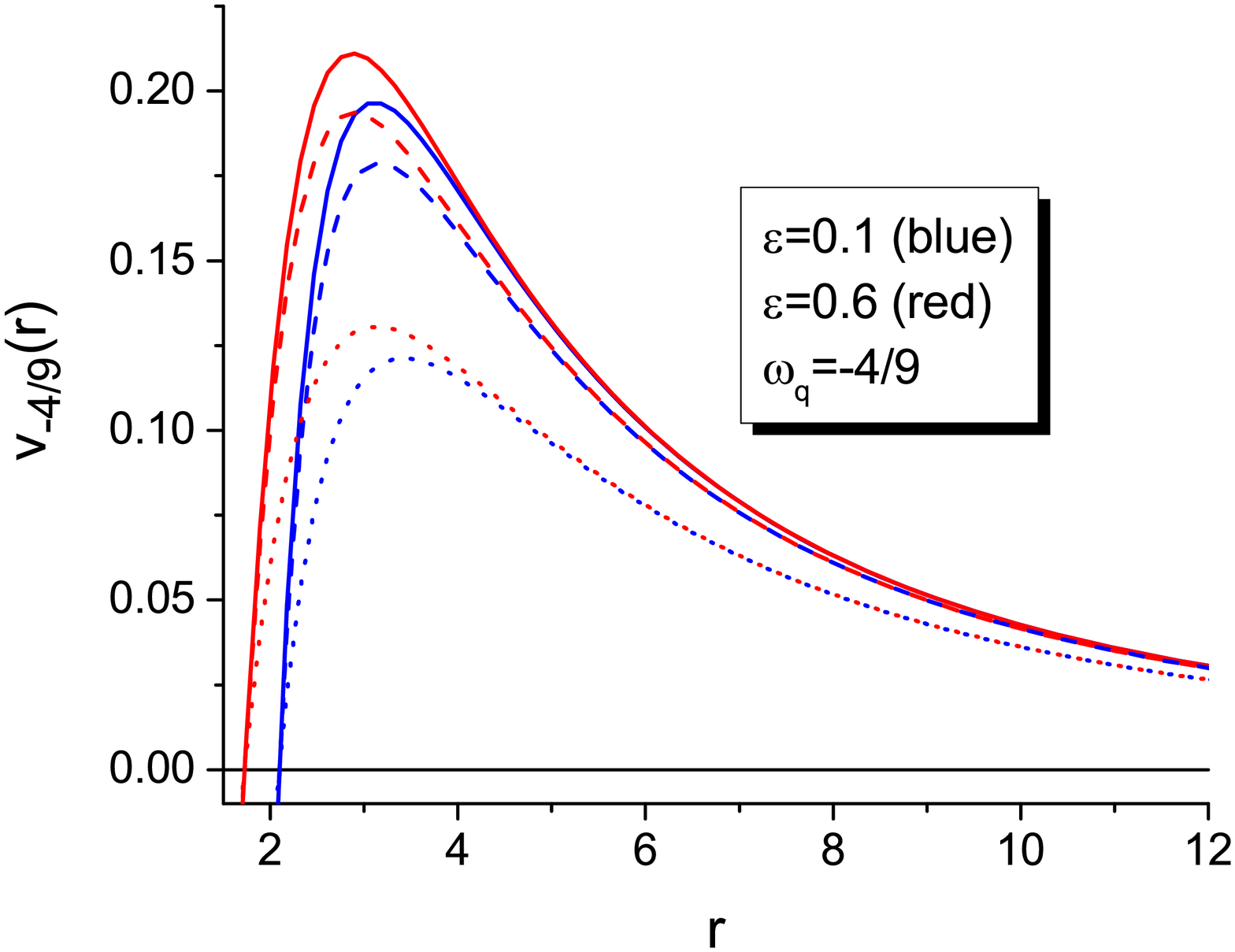}
\caption{The behavior of the effective potential of the scalar (solid), electromagnetic (dashed), and gravitational (dotted) perturbative fields with $\varepsilon^2$ for $c=0.05$, $l=2$  and $n=0$.}
\label{vsew}
\end{figure}

Fig. \ref{vsew} shows the dependence of the effective potential on the parameters $\varepsilon$. We can see when $\varepsilon$ is increased, the maximum potential increases, and its position moves toward the left. 

\begin{figure}[!h]
\centering
\includegraphics[scale=0.3]{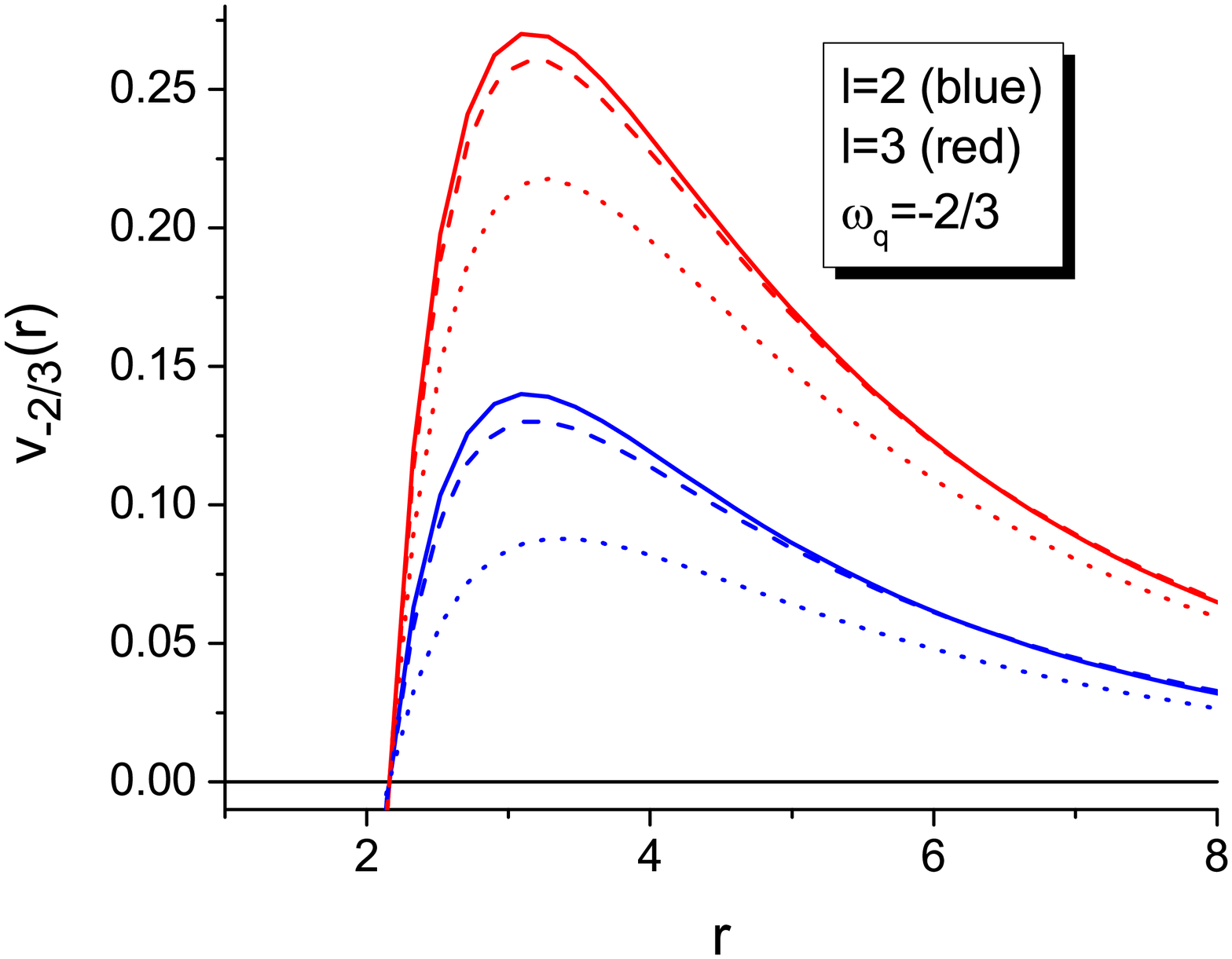}
\includegraphics[scale=0.3]{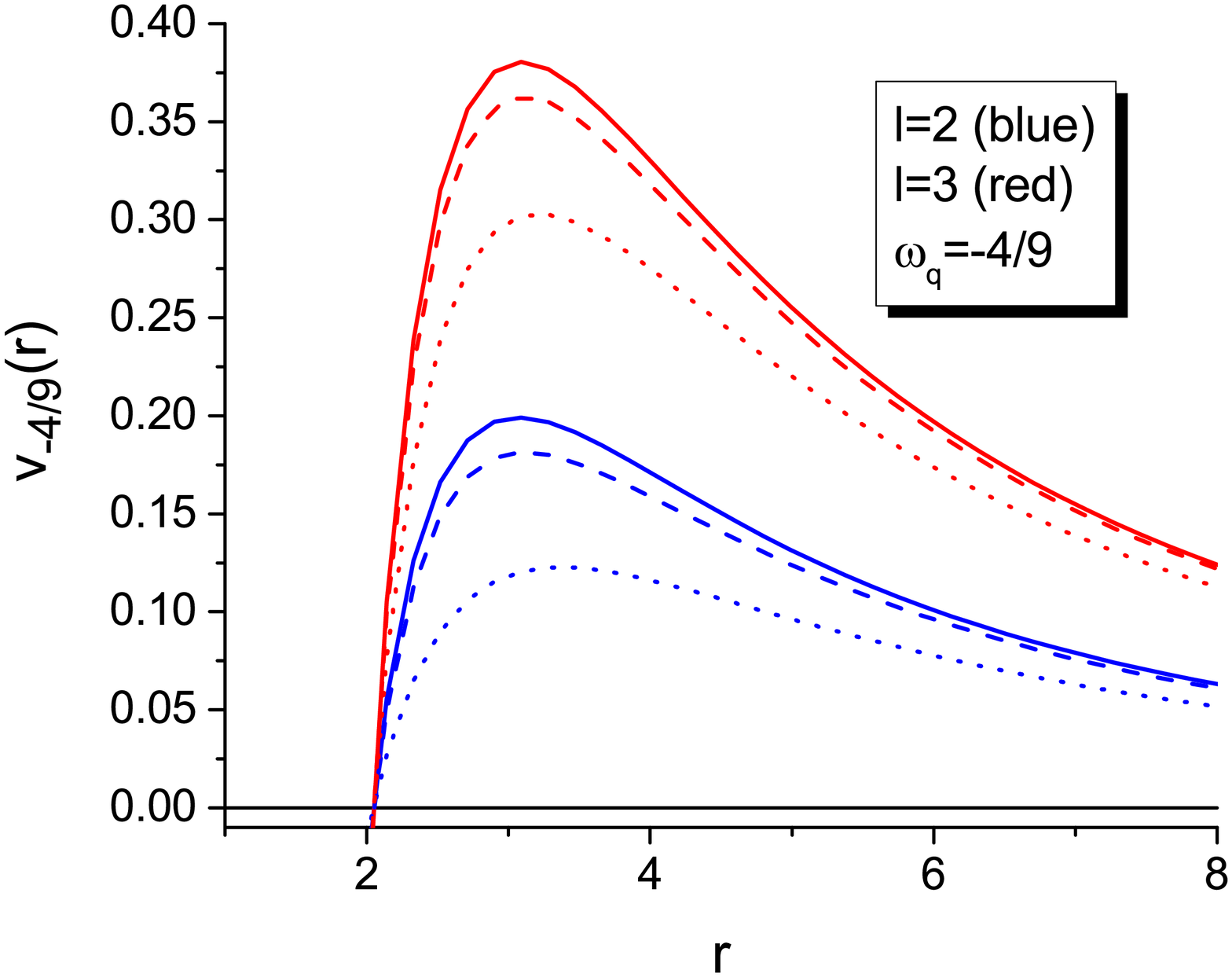}
\caption{The behavior of the effective potential of the scalar (solid), electromagnetic (dashed), and gravitational (dotted) perturbative fields with $l$ for $\varepsilon^2=0.2$, $c=0.05$ and $n=0$.}
\label{vslw}
\end{figure}

Finally Fig. \ref{vslw} shows the behavior of the effective potential on the angular harmonic index $l$. The maximum height of the potential increases and it moves toward the right as $l$ increases. On the other hand, the height of the potential increases with $\omega_q=-4/9$ ( $V_{\omega_q=-4/9} > V_{\omega_q=-2/3}$).

In general, the behavior for electromagnetic and gravitational perturbation is very similar to the scalar case. However, the maximum height of the potential is much higher for scalar perturbation and much lower for gravitational perturbation. When we consider $\omega_q=-4/9$, the behavior of $V_{\omega_q}(r)$ shows that differences occur on a smaller scale than with $\omega_q=-2/3$.

\section{Quasinormal modes of HBH--$\omega_q$}

For a BH, the QNM correspond to solutions of the wave equation given in Eq. (\ref{ecmcnsf}), which satisfy the appropriate boundary conditions. At the horizon, the boundary condition is such that the wave has to be purely ingoing and purely outgoing at spatial infinity. Only a discrete set of complex frequencies satisfies these conditions.    
    
To evaluate the QNM, we use here the third--order WKB approximation method developed by Schutz, Will \cite{Schutz:1985km} and Iyer \cite{Iyer:1986np}. The formula for the quasinormal frequencies is
\begin{equation}\label{ec.fmcn1}
\omega^2=\left[V_0+\left(-2V''_0\right)^{1/2}\Lambda\right]-i\left(n+\frac{1}{2}\right)\left(-2V''_0\right)^{1/2}\left[1+\Omega\right]\,,
\end{equation}
where
\begin{eqnarray}
\Lambda&=&\frac{1}{\left(-2V''_0\right)^{1/2}}\left\{
\frac{1}{8}\left(\frac{V_0^{(4)}}{V''_0}\right)\left(\frac{1}{4}+\alpha^2\right)
-\frac{1}{288}\left(\frac{{V'''}_0}{V''_0}\right)^2\left(7+60\alpha^2\right)
\right\}\,,\label{ecu.wkbl}\\
\Omega&=&\frac{1}{-2V''_0}
\left\{
\frac{5}{6912}\left(\frac{V'''_0}{V''_0}\right)^4\left(77+188\alpha^2\right)
-
\frac{1}{384}\left(\frac{{V'''}_0^2V_0^{(4)}}{{V''}_0^3}\right)\left(51+100\alpha^2\right)
\right.\nonumber\\
&&
-\left.
\frac{1}{288}\left(\frac{V_0^{(6)}}{V''_0}\right)\left(5+4\alpha^2\right)
+
\frac{1}{288}\left(\frac{{V'''}_0V_0^{(5)}}{{V''}_0^2}\right)\left(19+28\alpha^2\right)
+
\frac{1}{2304}\left(\frac{V_0^{(4)}}{V''_0}\right)^2\left(67+68\alpha^2\right)
\right\}\label{ecu.wkbo}\,,
\end{eqnarray}
with
\begin{equation}
\alpha=n+\frac{1}{2}
,\quad V_0^{(n)}=\frac{d^nV_{\omega_q}}{dr_*^n}\Big|_{r_*=r_*(r_p)}
\,,
\end{equation}
where $r_*(r_p)$ indicates the value of the variable $r_*$ at which the effective potential obtains its maximum.  

In Table \ref{scvsc}--\ref{gravsep} we show the spectrum for the frequencies of the QNM of the scalar, electromagnetic and gravitational perturbations for $\omega_q=-2/3$ and $\omega_q=-4/9$, respectively. It is worth mentioning that WKB method works best for $l>n$, while for $l=n$, it does not provide satisfactory results, as other authors have shown (see \cite{Konoplya:2003ii}). 

\begin{table}[th]
\begin{center}
\begin{tabular}{c c c c c c c c c }
\hline
\hline
\multicolumn{9}{ c }{Scalar perturbations}\\ \hline
&&\multicolumn{3}{ c}{$\omega_q=-2/3$}&\multicolumn{4}{ c}{$\omega_q=-4/9$}\\\hline 
$n$&$l$&$c=0.01$&$c=0.05$&$c=0.1$&&$c=0.01$&$c=0.05$&$c=0.1$\\
0&0&0.0955-0.1035$i$	
&0.0663-0.0837$i$	
&0.0256-0.0475$i$
&$\quad$	
&0.0994-0.1050$i$
&0.0873-0.0942$i$	
&0.0723-0.0806$i$\\
&1&
0.2804-0.0887$i$	
&0.2160-0.0667$i$	
&0.1165-0.0332$i$	
&
&0.2885-0.0912$i$
&0.2609-0.0809$i$	
&0.2265-0.0680$i$\\
&2&
0.4674-0.0879$i$	
&0.3658-0.0654$i$	
&0.2028-0.0324$i$	
&
&0.4797-0.0906$i$
&0.4353-0.0802$i$	
&0.3797-0.0674$i$\\ 
1&1&0.2538-0.2779$i$	
&0.1993-0.2072$i$	
&0.1107-0.1016$i$	
&
&0.2603-0.2862$i$
&0.2367-0.2531$i$	
&0.2069-0.2123$i$\\
&2&
0.4498-0.2681$i$	
&0.3543-0.1991$i$	
&0.1990-0.0979$i$	
&
&0.4613-0.2764$i$
&0.4194-0.2445$i$	
&0.3669-0.2053$i$\\
&3&
0.6411-0.2653$i$	
&0.5057-0.1968$i$	
&0.2843-0.0971$i$	
&
&0.6572-0.2737$i$
&0.5976-0.2422$i$	
&0.5227-0.2033$i$\\ 
2&2&0.4214-0.4557$i$	
&0.3355-0.3375$i$	
&0.1923-0.1648$i$	
&
&0.4315-0.4700$i$
&0.3937-0.4156$i$	
&0.3459-0.3486$i$\\
&3&
0.6187-0.4478$i$	
&0.4907-0.3315$i$	
&0.2792-0.1626$i$	
&
&0.6338-0.4620$i$
&0.5773-0.4086$i$	
&0.5062-0.3429$i$\\
&4&
0.8120-0.4441$i$	
&0.6431-0.3288$i$	
&0.3645-0.1617$i$	
&
&0.8319-0.4582$i$
&0.7573-0.4053$i$	
&0.6634-0.3402$i$\\ 
\hline
\hline		\end{tabular}
\caption{Quasinormal frequencies for the scalar perturbations for several values of the parameter $c$, with $\epsilon^2=0.2$.}
\label{scvsc}
\end{center}
\end{table}

\begin{table}[t]
\begin{center}
\begin{tabular}{c c c c c c c c c}
\hline
\hline
\multicolumn{9}{ c }{Scalar perturbations}\\ \hline
&&\multicolumn{3}{ c}{$\omega_q=-2/3$}&\multicolumn{4}{ c}{$\omega_q=-4/9$}\\\hline 
$n$&$l$&$\epsilon^2=0.1$&$\epsilon^2=0.4$&$\epsilon^2=0.7$&$\quad$&$\epsilon^2=0.1$&$\epsilon^2=0.4$&$\epsilon^2=0.7$\\
0&0&0.0665-0.0857$i$	
&0.0637-0.0791$i$	
&0.0570-0.0749$i$
&$\quad$	
&0.0882-0.0969$i$
&0.0831-0.0885$i$	
&0.0755-0.0844$i$\\%
&1&
0.2141-0.0678$i$	
&0.2202-0.0642$i$	
&0.2267-0.0583$i$
&	
&0.2592-0.0823$i$
&0.2645-0.0773$i$	
&0.2695-0.0699$i$\\
&2&
0.3626-0.0663$i$	
&0.3728-0.0632$i$	
&0.3851-0.0578$i$	
&
&0.4323-0.0815$i$
&0.4418-0.0770$i$	
&0.4529-0.0698$i$\\ 
1&1&0.1974-0.2104$i$	
&0.2027-0.1992$i$	
&0.2032-0.1822$i$	
&
&0.2351-0.2578$i$
&0.2386-0.2419$i$	
&0.2345-0.2211$i$\\
&2&
0.3509-0.2019$i$	
&0.3615-0.1921$i$	
&0.3713-0.1755$i$	
&
&0.4161-0.2487$i$
&0.4259-0.2345$i$	
&0.4327-0.2126$i$\\
&3&
0.5010-0.1994$i$	
&0.5157-0.1901$i$	
&0.5319-0.1740$i$	
&
&0.5930-0.2460$i$
&0.6071-0.2326$i$	
&0.6208-0.2108$i$\\ 
2&2&0.3317-0.3423$i$	
&0.3424-0.3252$i$	
&0.3464-0.2976$i$	
&
&0.3901-0.4228$i$
&0.3992-0.3981$i$	
&0.3962-0.3627$i$\\
&3&
0.4856-0.3361$i$	
&0.5010-0.3199$i$	
&0.5137-0.2926$i$	
&
&0.5723-0.4154$i$
&0.5868-0.3919$i$	
&0.5942-0.3556$i$\\
&4&
0.6368-0.3332$i$	
&0.6563-0.3176$i$	
&0.6757-0.2906$i$	
&
&0.7511-0.4119$i$
&0.7698-0.3891$i$	
&0.7849-0.3529$i$\\ 
\hline
\hline		\end{tabular}
\caption{Quasinormal frequencies for the scalar perturbations for several values of the parameter $\epsilon^2$, with $c=0.05$.}
\label{scvsep}
\end{center}
\end{table}

\begin{table}[th]
	\begin{center}
		\begin{tabular}{c c c c c c c c c }
			\hline
			\hline
			\multicolumn{9}{ c }{Electromagnetic perturbations}\\ \hline
			&&\multicolumn{3}{ c}{$\omega_q=-2/3$}&\multicolumn{4}{ c}{$\omega_q=-4/9$}\\\hline 
			$n$&$l$&$c=0.01$&$c=0.05$&$c=0.1$&&$c=0.01$&$c=0.05$&$c=0.1$\\
			0&1&
			0.2407-0.0839$i$	
			&0.1929-0.0627$i$	
			&0.1113-0.0314$i$
			&$\quad$	
			&0.2462-0.0865$i$
			&0.2253-0.0769$i$	
			&0.1987-0.0648$i$\\
			&2&
			0.4444-0.0862$i$	
			&0.3526-0.0640$i$	
			&0.1998-0.0318$i$	
			&
			&0.4552-0.0889$i$
			&0.4147-0.0788$i$	
			&0.3636-0.0663$i$\\
			&3&
			0.6376-0.0868$i$	
			&0.5047-0.0644$i$	
			&0.2849-0.0319$i$	
			&
			&0.6532-0.0896$i$
			&0.5945-0.0793$i$	
			&0.5206-0.0667$i$\\ 
			1&1&0.2093-0.2664$i$	
			&0.1721-0.1969$i$	
			&0.1043-0.0964$i$	
			&
			&0.2134-0.2749$i$
			&0.1972-0.2433$i$	
			&0.1761-0.2044$i$\\
			&2&
			0.4259-0.2633$i$	
			&0.3403-0.1950$i$	
			&0.1958-0.0962$i$	
			&
			&0.4359-0.2717$i$
			&0.3980-0.2406$i$	
			&0.3501-0.2021$i$\\
			&3&
			0.6245-0.2628$i$	
			&0.4960-0.1947$i$	
			&0.2822-0.0962$i$	
			&
			&0.6395-0.2712$i$
			&0.5827-0.2401$i$	
			&0.5111-0.2017$i$\\ 
			2&2&0.3961-0.4482$i$	
			&0.3200-0.3311$i$	
			&0.1888-0.1618$i$	
			&
			&0.4048-0.4626$i$
			&0.3712-0.4093$i$	
			&0.3282-0.3436$i$\\
			&3&
			0.6015-0.4438$i$	
			&0.4805-0.3281$i$	
			&0.2769-0.1611$i$	
			&
			&0.6155-0.4580$i$
			&0.5619-0.4053$i$	
			&0.4942-0.3402$i$\\
			&4&
			0.7989-0.4416$i$	
			&0.6354-0.3267$i$	
			&0.3628-0.1609$i$	
			&
			&0.8180-0.4557$i$
			&0.7456-0.4032$i$	
			&0.6543-0.3385$i$\\ 
			\hline
			\hline		\end{tabular}
		\caption{Quasinormal frequencies for the electromagnetic perturbations for several values of the parameter $c$, with $\epsilon^2=0.2$.}
		\label{elvsc}
	\end{center}
\end{table}

\begin{table}[t]
	\begin{center}
		\begin{tabular}{c c c c c c c c c}
			\hline
			\hline
			\multicolumn{9}{ c }{Electromagnetic perturbations}\\ \hline
			&&\multicolumn{3}{ c}{$\omega_q=-2/3$}&\multicolumn{4}{ c}{$\omega_q=-4/9$}\\\hline 
			$n$&$l$&$\epsilon^2=0.1$&$\epsilon^2=0.4$&$\epsilon^2=0.7$&$\quad$&$\epsilon^2=0.1$&$\epsilon^2=0.4$&$\epsilon^2=0.7$\\
			0&1&0.1908-0.0636$i$	
			&0.1975-0.0602$i$	
			&0.2050-0.0541$i$
			&$\quad$	
			&0.2232-0.0783$i$
			&0.2298-0.0732$i$	
			&0.2367-0.0645$i$\\%
			&2&
			0.3494-0.0649$i$	
			&0.3597-0.0618$i$	
			&0.3725-0.0563$i$
			&	
			&0.4115-0.0801$i$
			&0.4216-0.0756$i$	
			&0.4338-0.0680$i$\\
			&3&
			0.5003-0.0652$i$	
			&0.5145-0.0623$i$	
			&0.5324-0.0569$i$	
			&
			&0.5901-0.0806$i$
			&0.6040-0.0762$i$	
			&0.6211-0.0688$i$\\ 
			1&1&0.1696-0.2000$i$	
			&0.1765-0.1890$i$	
			&0.1782-0.1704$i$	
			&
			&0.1949-0.2480$i$
			&0.2006-0.2317$i$	
			&0.1971-0.2070$i$\\
			&2&
			0.3367-0.1977$i$	
			&0.3478-0.1880$i$	
			&0.3584-0.1711$i$	
			&
			&0.3944-0.2447$i$
			&0.4053-0.2303$i$	
			&0.4130-0.2072$i$\\
			&3&
			0.4913-0.1973$i$	
			&0.5063-0.1880$i$	
			&0.5228-0.1718$i$	
			&
			&0.5780-0.2440$i$
			&0.5926-0.2304$i$	
			&0.6070-0.2081$i$\\ 
			2&2&0.3159-0.3358$i$	
			&0.3277-0.3189$i$	
			&0.3327-0.2904$i$	
			&
			&0.3671-0.4165$i$
			&0.3777-0.3915$i$	
			&0.3756-0.3539$i$\\
			&3&
			0.4753-0.3326$i$	
			&0.4911-0.3166$i$	
			&0.5044-0.2889$i$	
			&
			&0.5567-0.4121$i$
			&0.5720-0.3883$i$	
			&0.5802-0.3510$i$\\
			&4&
			0.6291-0.3310$i$	
			&0.6488-0.3155$i$	
			&0.6686-0.2884$i$	
			&
			&0.7393-0.4098$i$
			&0.7585-0.3869$i$	
			&0.7741-0.3501$i$\\ 
			\hline
			\hline		\end{tabular}
		\caption{Quasinormal frequencies for the electromagnetic perturbations for several values of the parameter $\epsilon^2$, with $c=0.05$.}
		\label{elvsep}
	\end{center}
\end{table}

\begin{table}[th]
	\begin{center}
		\begin{tabular}{c c c c c c c c c }
			\hline
			\hline
			\multicolumn{9}{ c }{Gravitational perturbations}\\ \hline
			&&\multicolumn{3}{ c}{$\omega_q=-2/3$}&\multicolumn{4}{ c}{$\omega_q=-4/9$}\\\hline 
			$n$&$l$&$c=0.01$&$c=0.05$&$c=0.1$&&$c=0.01$&$c=0.05$&$c=0.1$\\
			0&2&
			0.3622-0.0809$i$	
			&0.2868-0.0611$i$	
			&0.1622-0.0312$i$
			&$\quad$	
			&0.3712-0.0832$i$
			&0.3379-0.0741$i$	
			&0.2960-0.0627$i$\\
			&3&
			0.5817-0.0843$i$	
			&0.4603-0.0631$i$	
			&0.2598-0.0317$i$	
			&
			&0.5959-0.0869$i$
			&0.5423-0.0771$i$	
			&0.4749-0.0650$i$\\
			&4&
			0.7850-0.0856$i$	
			&0.6209-0.0638$i$	
			&0.3500-0.0318$i$	
			&
			&0.8043-0.0883$i$
			&0.7317-0.0783$i$	
			&0.6405-0.0659$i$\\ 
			1&2&0.3385-0.2488$i$	
			&0.2717-0.1872$i$	
			&0.1575-0.0945$i$	
			&
			&0.3462-0.2561$i$
			&0.3167-0.2277$i$	
			&0.2791-0.1923$i$\\
			&3&
			0.5673-0.2557$i$	
			&0.4509-0.1908$i$	
			&0.2568-0.0954$i$	
			&
			&0.5809-0.2635$i$
			&0.5293-0.2338$i$	
			&0.4644-0.1969$i$\\
			&4&
			0.7743-0.2585$i$	
			&0.6138-0.1923$i$	
			&0.3478-0.0957$i$	
			&
			&0.7931-0.2665$i$
			&0.7221-0.2362$i$	
			&.6327-0.1987$i$\\ 
			2&2&0.3003-0.4260$i$	
			&0.2473-0.3189$i$	
			&0.1495-0.1593$i$	
			&
			&0.3059-0.4389$i$
			&0.2825-0.3897$i$	
			&0.2518-0.3287$i$\\
			&3&
			0.5421-0.4324$i$	
			&0.4341-0.3220$i$	
			&0.2513-0.1599$i$	
			&
			&0.5545-0.4458$i$
			&0.5066-0.3953$i$	
			&0.4459-0.3328$i$\\
			&4&
			0.7547-0.4348$i$	
			&0.6008-0.3230$i$	
			&0.3435-0.1601$i$	
			&
			&0.7726-0.4484$i$
			&0.7044-0.3973$i$	
			&0.6183-0.3341$i$\\ 
			\hline
			\hline		\end{tabular}
		\caption{Quasinormal frequencies for the gravitational perturbations for several values of the parameter $c$, with $\epsilon^2=0.2$.}
		\label{gravsc}
	\end{center}
\end{table}

\begin{table}[t]
	\begin{center}
		\begin{tabular}{c c c c c c c c c}
			\hline
			\hline
			\multicolumn{9}{ c }{Gravitational perturbations}\\ \hline
			&&\multicolumn{3}{ c}{$\omega_q=-2/3$}&\multicolumn{4}{ c}{$\omega_q=-4/9$}\\\hline 
			$n$&$l$&$\epsilon^2=0.1$&$\epsilon^2=0.4$&$\epsilon^2=0.7$&$\quad$&$\epsilon^2=0.1$&$\epsilon^2=0.4$&$\epsilon^2=0.7$\\
			0&2&0.2841-0.0620$i$	
			&0.2925-0.0589$i$	
			&0.3026-0.0535$i$
			&$\quad$	
			&0.3354-0.0754$i$
			&0.3433-0.0709$i$	
			&0.3526-0.0635$i$\\%
			&3&
			0.4563-0.0639$i$	
			&0.4693-0.0610$i$	
			&0.4855-0.0558$i$
			&	
			&0.5384-0.0783$i$
			&0.5510-0.0741$i$	
			&0.5664-0.0670$i$\\
			&4&
			0.6155-0.0646$i$	
			&0.6328-0.0617$i$	
			&0.6545-0.0565$i$	
			&
			&0.7265-0.0795$i$
			&0.7432-0.0753$i$	
			&0.7641-0.0682$i$\\ 
			1&2&0.2687-0.1899$i$	
			&0.2779-0.1801$i$	
			&0.2855-0.1633$i$	
			&
			&0.3137-0.2319$i$
			&0.3223-0.2174$i$	
			&0.3269-0.1946$i$\\
			&3&
			0.4465-0.1934$i$	
			&0.4603-0.1843$i$	
			&0.4752-0.1684$i$	
			&
			&0.5250-0.2375$i$
			&0.5385-0.2243$i$	
			&0.5513-0.2027$i$\\
			&4&
			0.6082-0.1948$i$	
			&0.6261-0.1859$i$	
			&0.6469-0.1702$i$	
			&
			&0.7165-0.2399$i$
			&0.7340-0.2270$i$	
			&0.7529-0.2055$i$\\ 
			2&2&0.2438-0.3237$i$	
			&0.2535-0.3067$i$	
			&0.2548-0.2789$i$	
			&
			&0.2790-0.3969$i$
			&0.2872-0.3719$i$	
			&0.2807-0.3354$i$\\
			&3&
			0.4292-0.3265$i$	
			&0.4440-0.3107$i$	
			&0.4556-0.2835$i$	
			&
			&0.5016-0.4019$i$
			&0.5160-0.3788$i$	
			&0.5227-0.3425$i$\\
			&4&
			0.5947-0.3274$i$	
			&0.6135-0.3120$i$	
			&0.6321-0.2853$i$	
			&
			&0.6983-0.4037$i$
			&0.7168-0.3813$i$	
			&0.7313-0.3451$i$\\ 
			\hline
			\hline		\end{tabular}
		\caption{Quasinormal frequencies for the gravitational perturbations for several values of the parameter $\epsilon^2$, with $c=0.05$.}
		\label{gravsep}
	\end{center}
\end{table}

\begin{figure}[!h]
\centering
\includegraphics[scale=0.3]{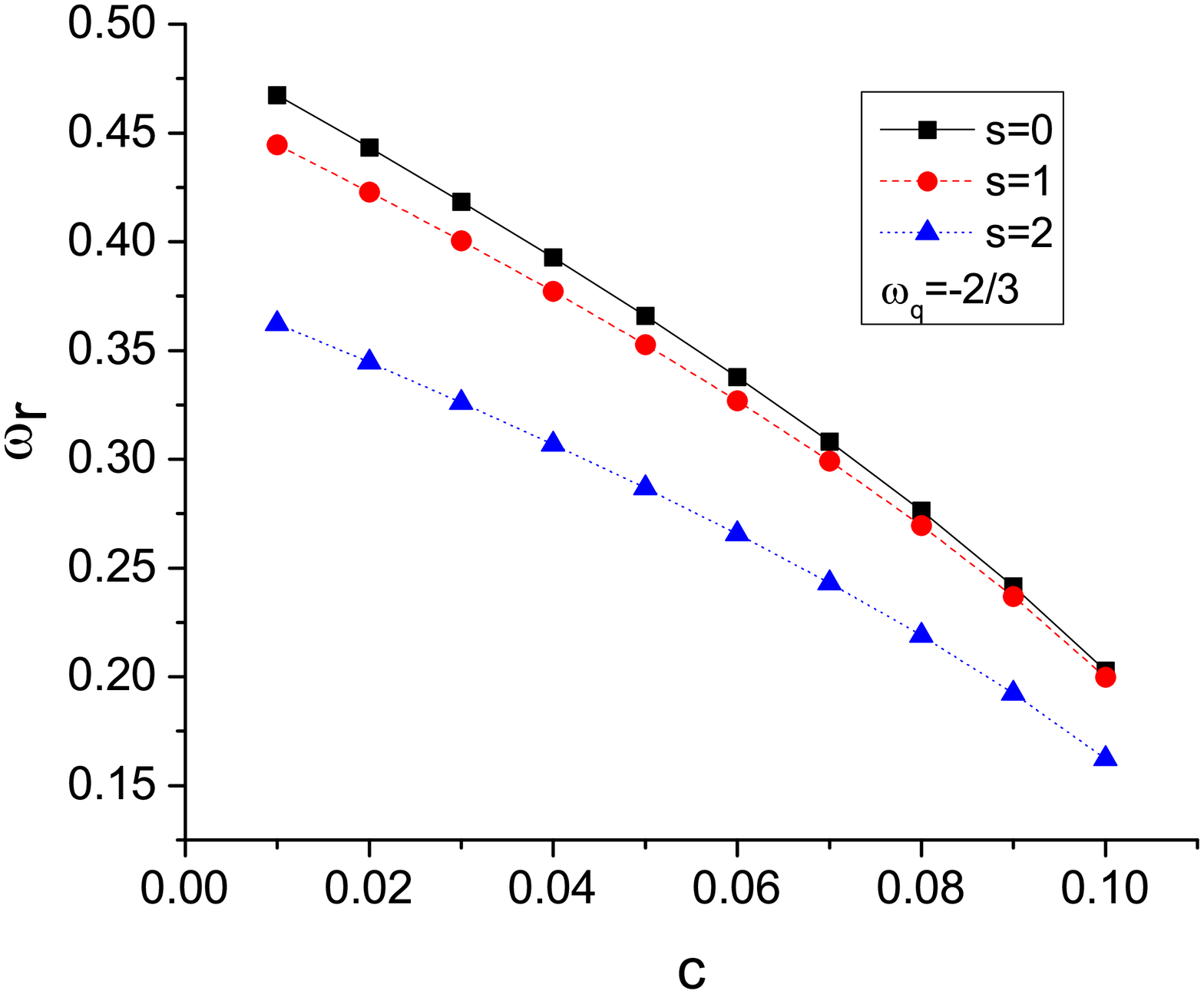}
\includegraphics[scale=0.3]{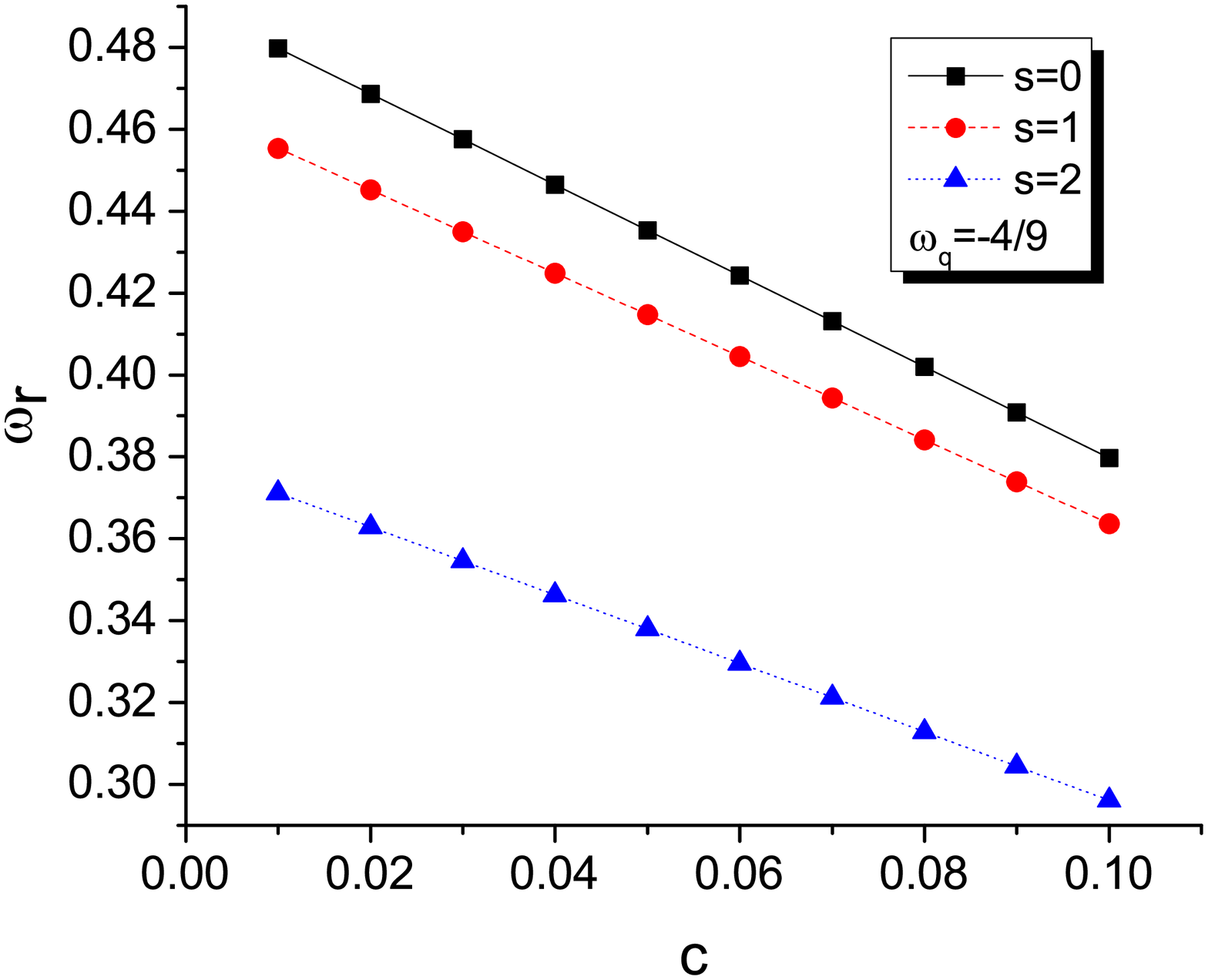}
\caption{Quasinormal frequencies for a massless scalar field on Hayward BH--$\omega_q$  for $\varepsilon^2=0.2$, $n=0$ and $l=2$.}
\label{fmrvsc}
\end{figure}

\begin{figure}[!h]
	\centering
	\includegraphics[scale=0.3]{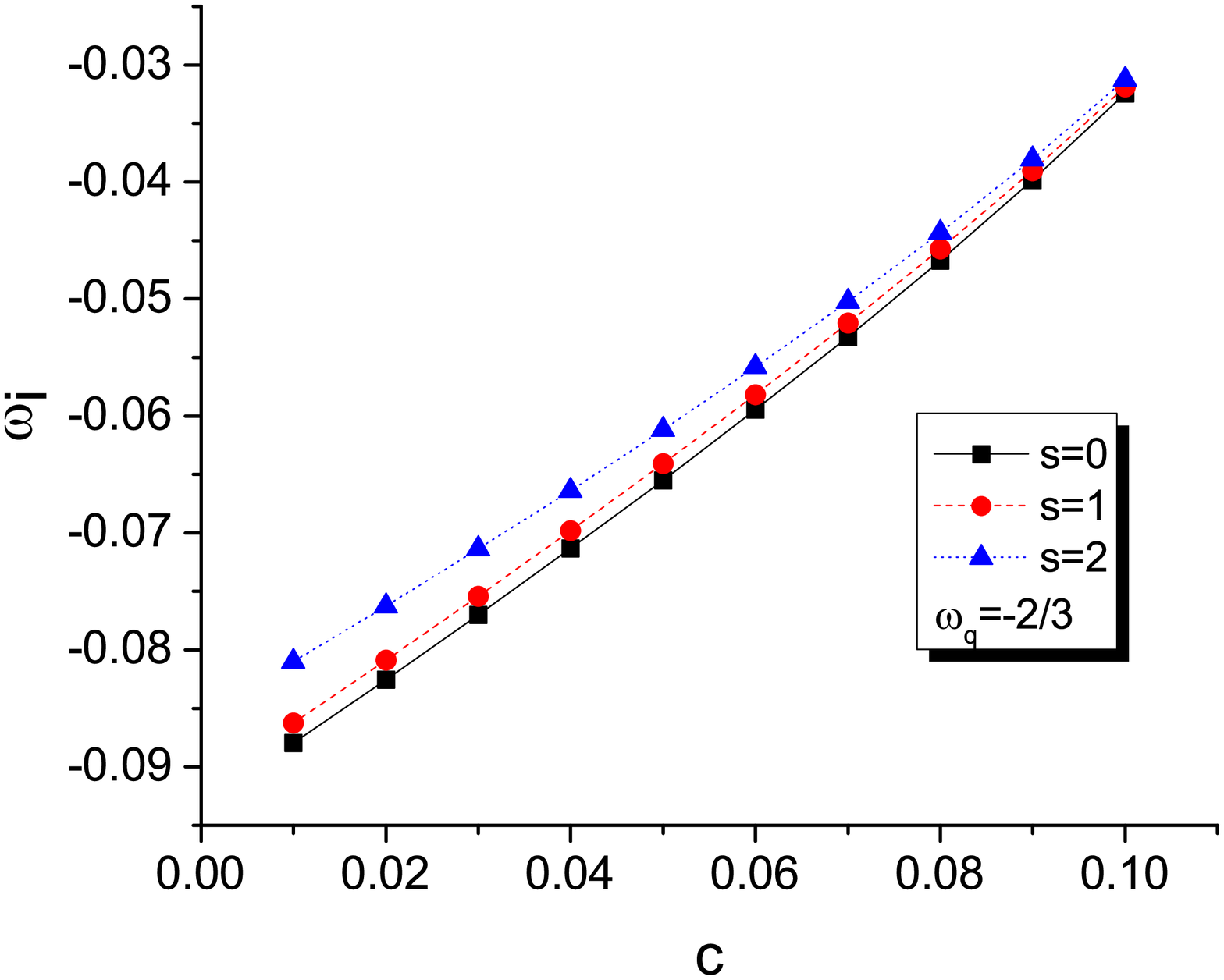}
	\includegraphics[scale=0.3]{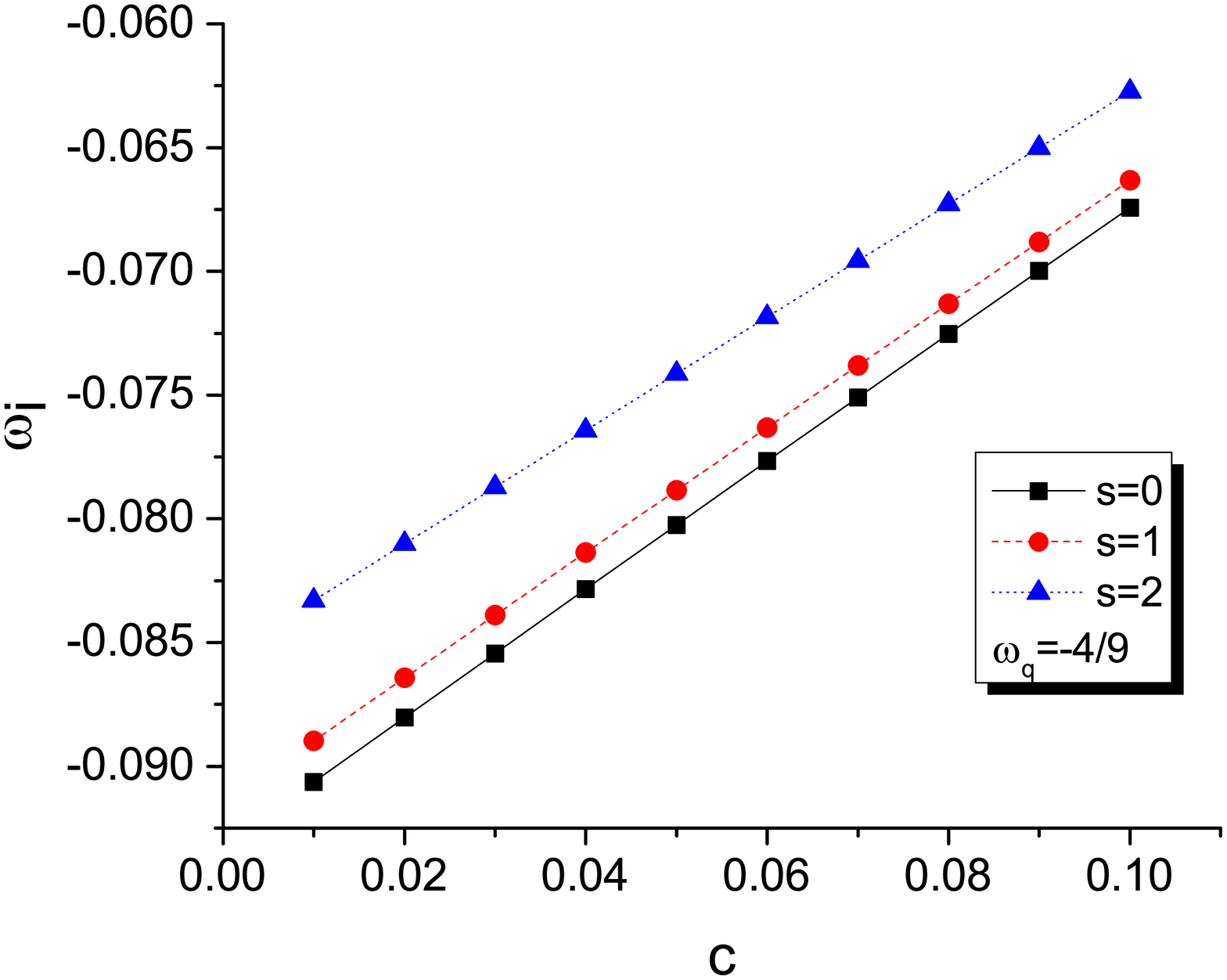}
	\caption{Quasinormal frequencies for a massless scalar field on Hayward BH-$\omega_q$  for $\varepsilon^2=0.2$, $n=0$ and $l=2$.}
	\label{fmivsc}
\end{figure}

\begin{figure}[!h]
\centering
\includegraphics[scale=0.3]{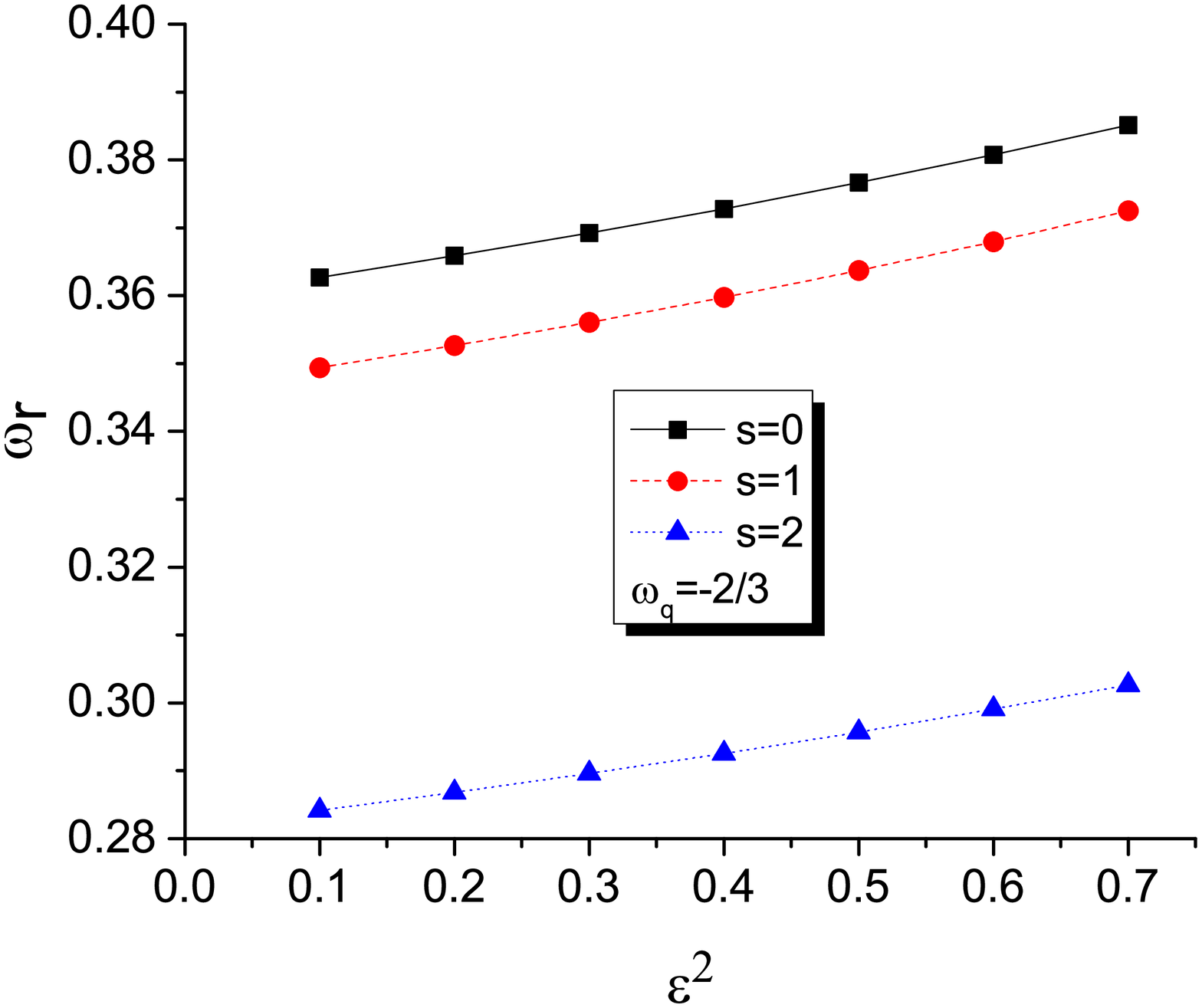}
\includegraphics[scale=0.3]{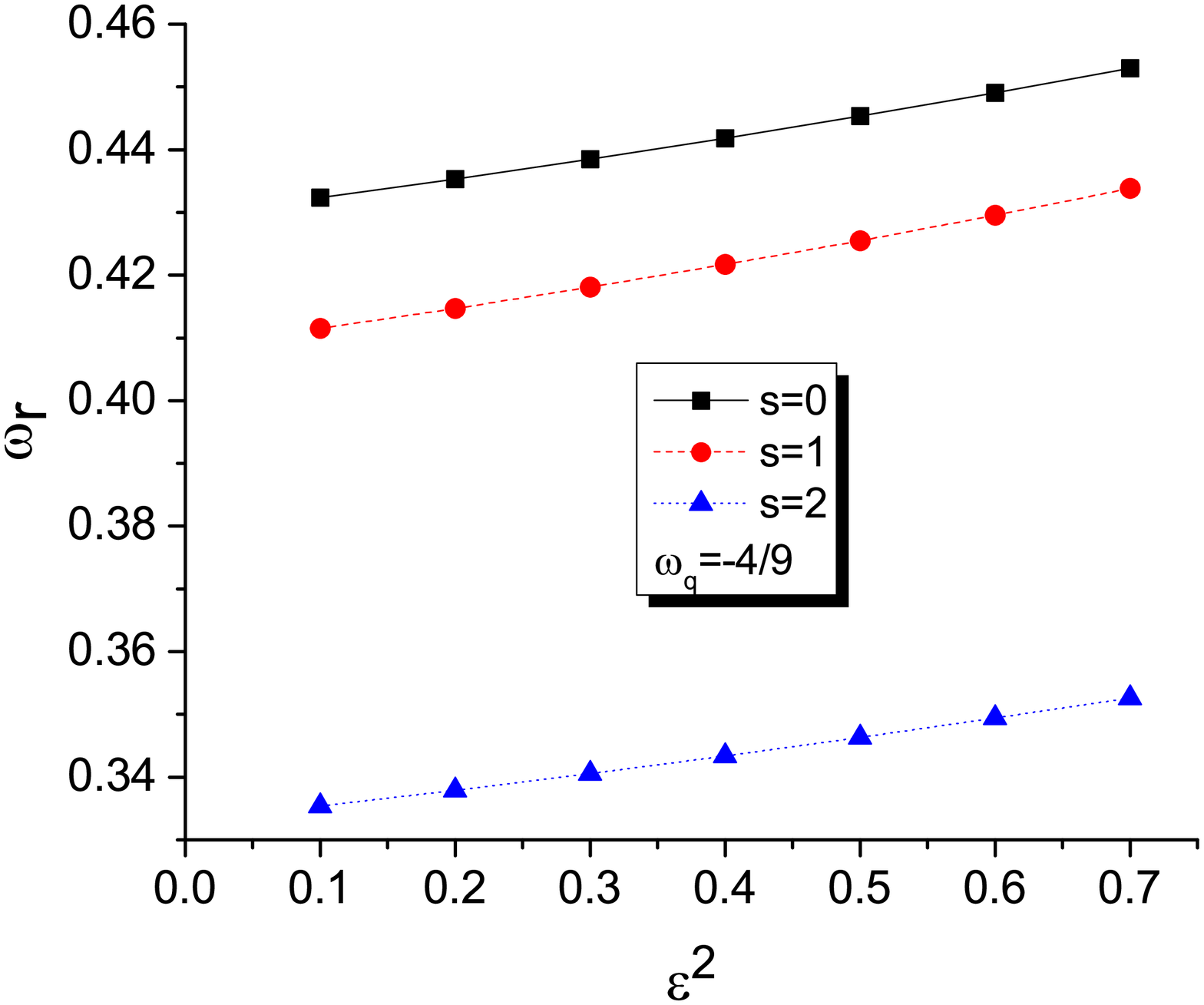}
\caption{Quasinormal frequencies for a massless scalar field on Hayward BH--$\omega_q$  for $c=0.05$, $n=0$ and $l=2$.}
\label{fmrvse}
\end{figure}

\begin{figure}[!h]
	\centering
	\includegraphics[scale=0.3]{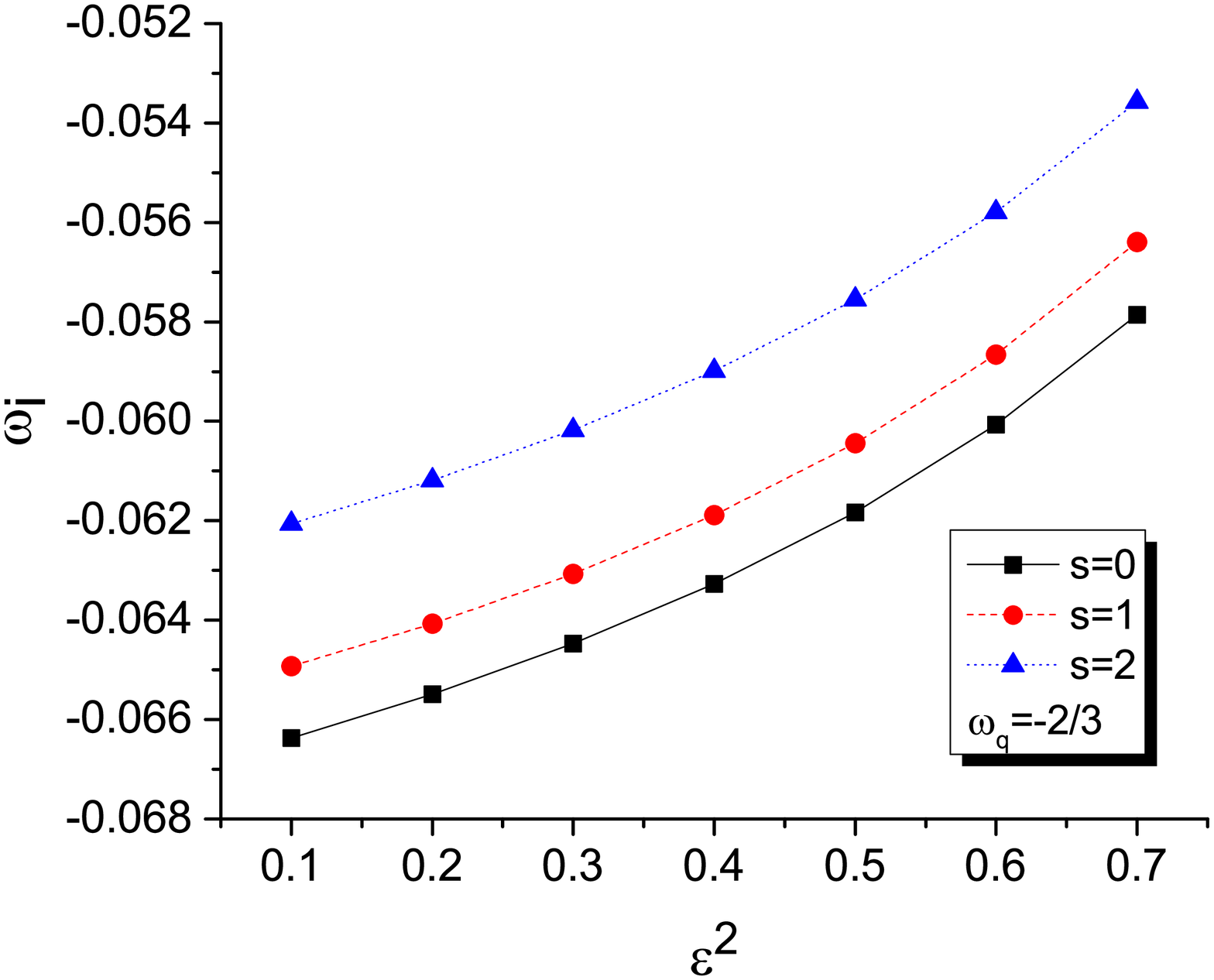}
	\includegraphics[scale=0.3]{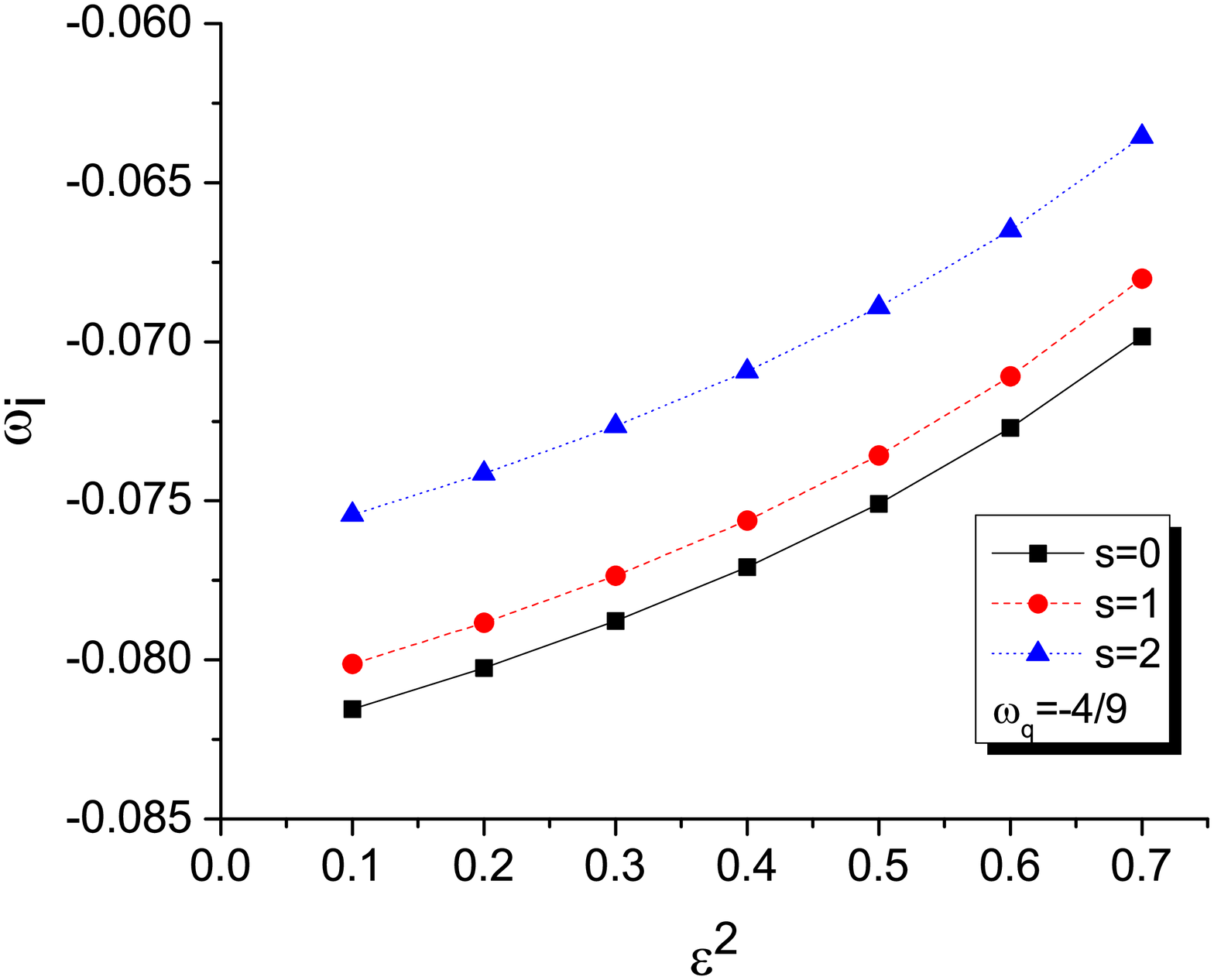}
	\caption{Quasinormal frequencies for a massless scalar field on Hayward BH--$\omega_q$  for $c=0.05$, $n=0$ and $l=2$.}
	\label{fmivse}
\end{figure}

The complex quasinormal frequencies are shown in Figs. \ref{fmrvsc}--\ref{fmivse}. In Fig. \ref{fmrvsc} we can see that the real value of the QNM frequency $\omega_r$ decreases when $c$ increases for both values of the quintessence state parameter, $\omega_q=-2/3$, and $\omega_q=-4/9$, also it is possible mention that $\omega_{r_{sc}}>\omega_{r_{elec}}>\omega_{r_{grav}}$. This means the perturbation in the fundamental mode ($n = 0$) with larger $c$ leads to a less intense QNM oscillation, i.e., the presence of quintessence suppress oscillation.

On the other hand, $\omega_i$ increase with increasing values of $c$, however, the increasing for the QNM frequencies for $\omega_q=-4/9$ is slower than for $\omega_q=-2/3$, for scalar electromagnetic and gravitational perturbations as we can see in Fig. \ref{fmivsc}

One of the important properties of the perturbations is the relaxation time, which is defined by the inverse of the imaginary part of QNM ($\tau=1/ |\omega_{i}|$), then we can mention that in presence of quintessence the relaxation time (damping rate) increases in all cases of perturbation.

In Figs. \ref{fmrvse} and \ref{fmivse} the real and imaginary parts of the QNM frequencies are plotted as function of the parameter $\varepsilon$ for scalar, electromagnetic and gravitational perturbations of the HBH--$\omega_q$. Here, $\omega_r$ and $\omega_i$ increase with the increase in $\epsilon$ for $\omega_q=-4/9$ and $\omega_q=-2/3$.  The relaxation times of the  perturbations for $\omega_q=-4/9$ and $\omega_q=-2/3$ qualitatively behave similarly, but quantitatively, their differences are $\tau_{sc}<\tau_{elec}<\tau_{grav}$. 

For the three types of perturbations studies here, the behaviors are very similar for the real and imaginary parts of the QNM frequencies when we varied $c$, keeping $\varepsilon$ and $\omega_q$ fixed and when we varied $\varepsilon$, keeping $c$ and $\omega_q$ fixed.

\section{Greybody Factor}

In this section we study the reflection and transmission coefficient $R(\omega)$ and $T(\omega)$ respectively for a scattering process of the scalar, electromagnetic and gravitational wave from BH--$\omega_q$. 

In this case, the incoming wave towards a BH--$\omega_q$ is partially transmitted and partially reflected by the potential barrier (\ref{ec.poseg}). So, the scattering behavior of the wave can be written in tortoise as;
 
\begin{eqnarray}
\xi(r_*)&=&T(\omega)e^{-i\omega r_*},\quad r_*\to -\infty\,,\\
\xi(r_*)&=&e^{-i\omega r_*}+R(\omega)e^{i\omega r_*},\quad r_*\to \infty\,,
\end{eqnarray}
where $R(\omega)$ and $T(\omega)$ are related by 
\begin{equation}\label{ec.cp}
|R(\omega)|^2+|T(\omega)|^2=1\,.
\end{equation}
So in general the greybody factor is defined as
\begin{equation}
\gamma_l=|T(\omega)|^2=1-|R(\omega)|^2\,.
\end{equation} 
In the WKB approximation, the reflection coefficient is given by
\begin{equation}\label{ecu.cr}
R(\omega)=\left(1+e^{-2\pi i\kappa}\right)^{-1/2}\,,
\end{equation} 
where to third-order in the WKB approximation, we have 
\begin{equation}
\kappa=i\frac{\omega^2-V_0}{\sqrt{-2V''_0}}-\Lambda+\left(n+\frac{1}{2}\right)\Omega\,,
\end{equation}
where $\Lambda$ and $\Omega$ are given by Eqs. (\ref{ecu.wkbl}) and (\ref{ecu.wkbo}), respectively. From Eq. (\ref{ecu.cr}), we can express the transmission coefficient as
\begin{equation}
\gamma_l=|T(\omega)|^2=1-\left|\left(1+e^{-2\pi i\kappa}\right)^{-1/2}\right|^2\,.
\end{equation} 
It is worth mentioning that if $\omega\ll V_0$, then the transmission coefficient is close to zero and the reflection coefficient is close to one, on the other hand, if $\omega\gg V_0$, then the reflection coefficient is close to zero, while the transmission coefficient is close to one, however, if $\omega\approx V_0$ the WKB approximation \cite{Konoplya:2009hv} has high accuracy. Therefore, we can find the transmission and reflection coefficients.

\begin{figure}[!h]
\centering
\includegraphics[scale=0.31]{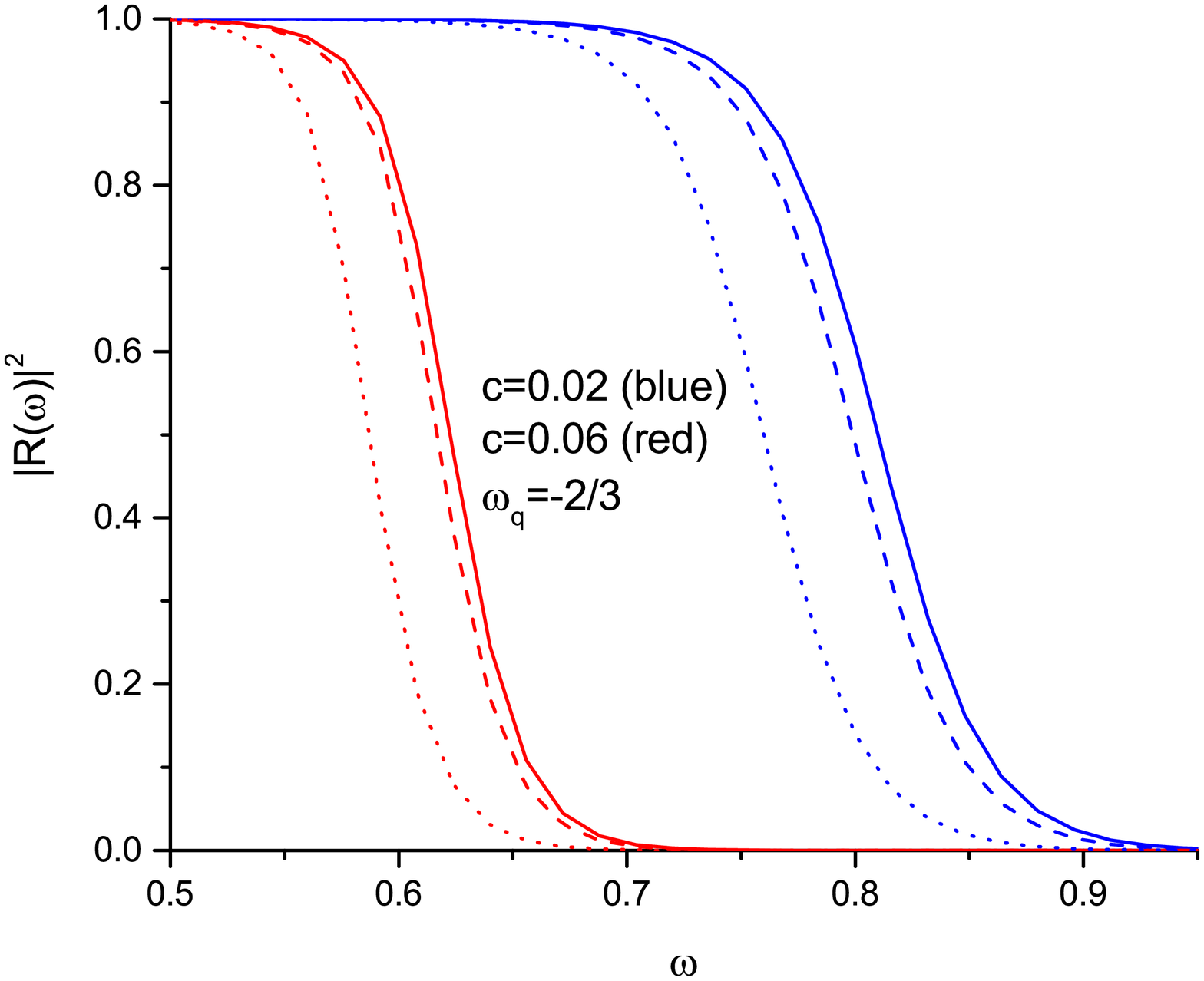}
\includegraphics[scale=0.31]{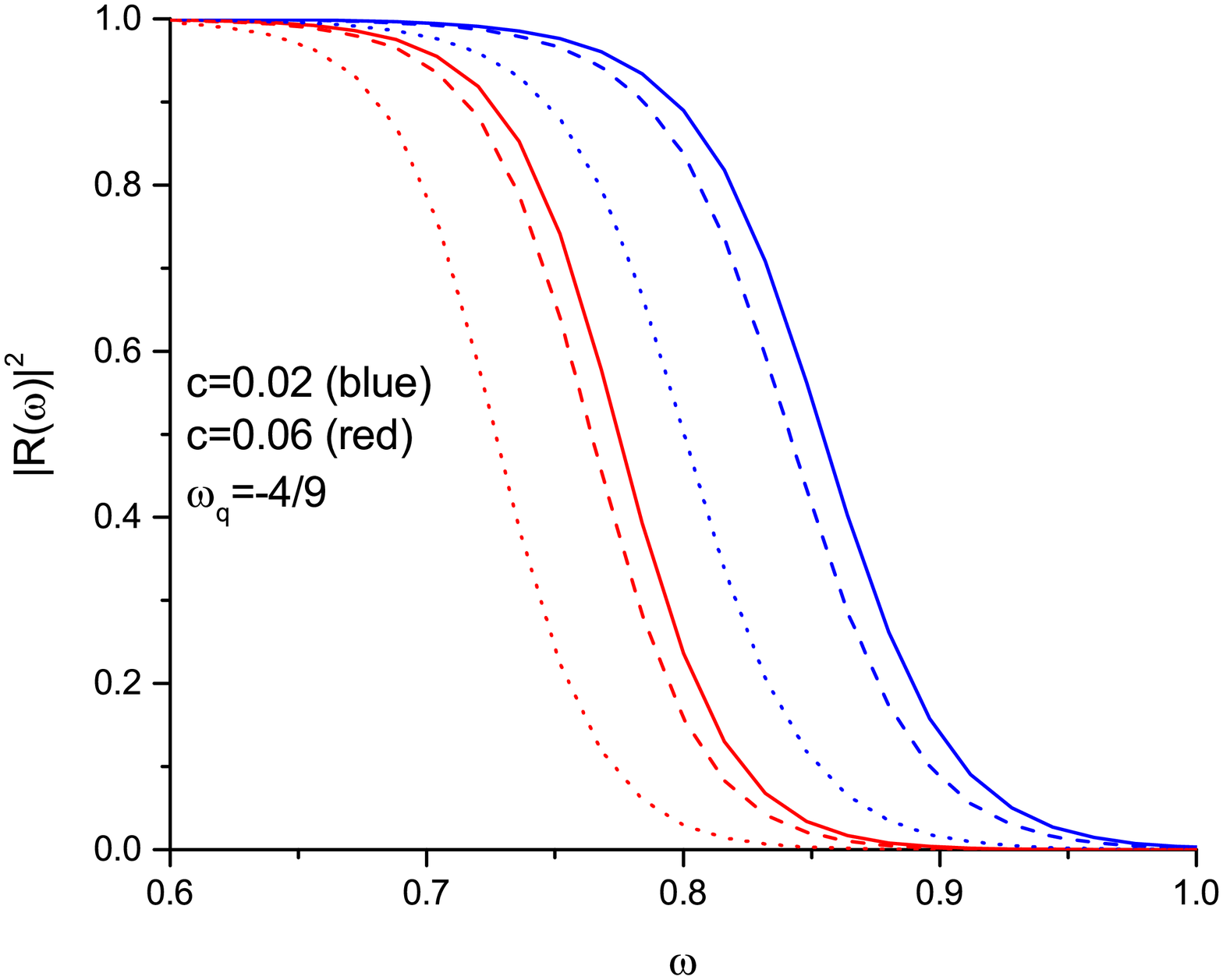}
\caption{The plot shows the reflection coefficients of the scattered scalar (solid), electromagnetic (dashed) and gravitational (dotted) wave for $l=4$, $\epsilon^2=0.2$ and $n=0$.}
\label{rc}
\end{figure}

\begin{figure}[!h]
\centering
\includegraphics[scale=0.31]{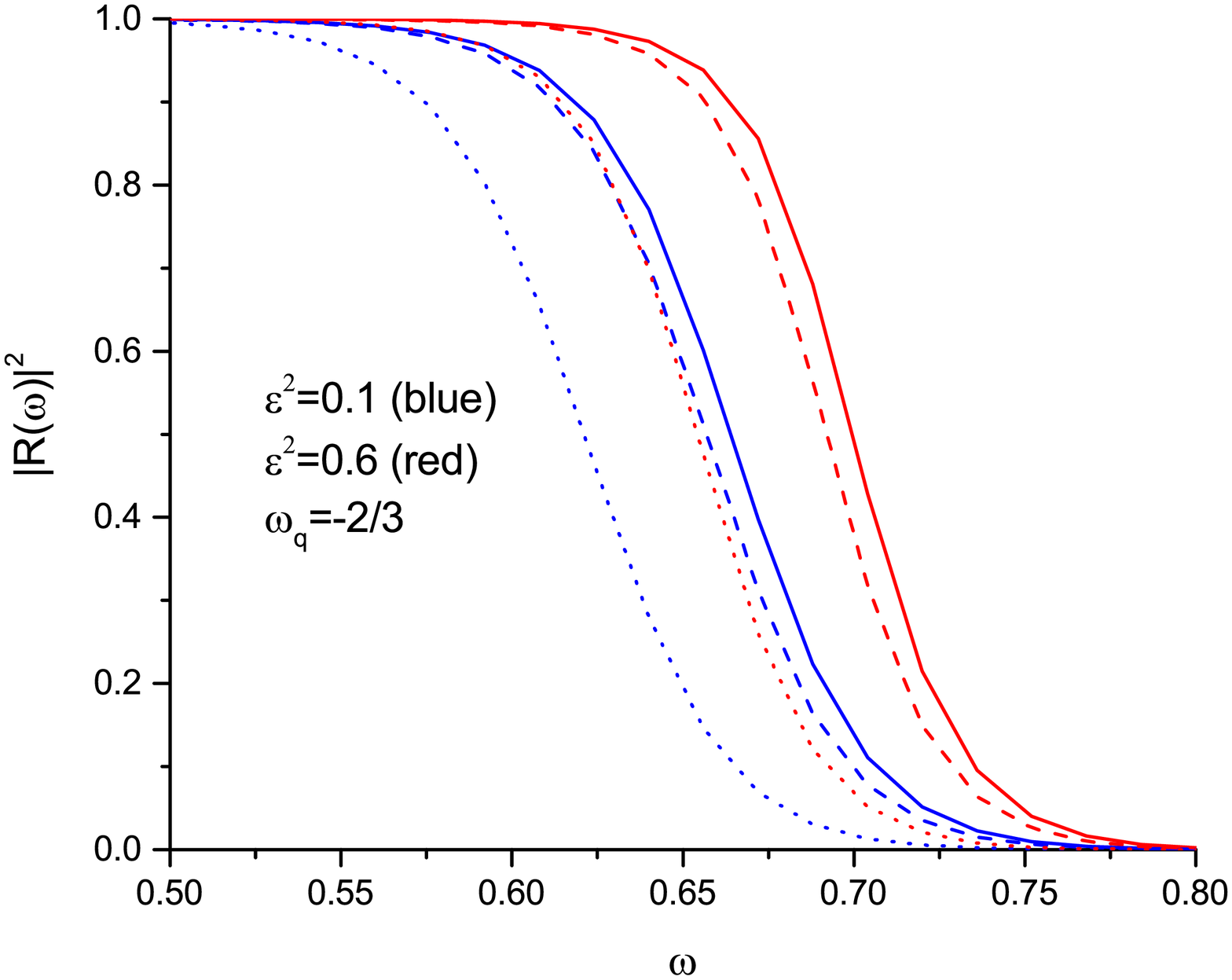}
\includegraphics[scale=0.31]{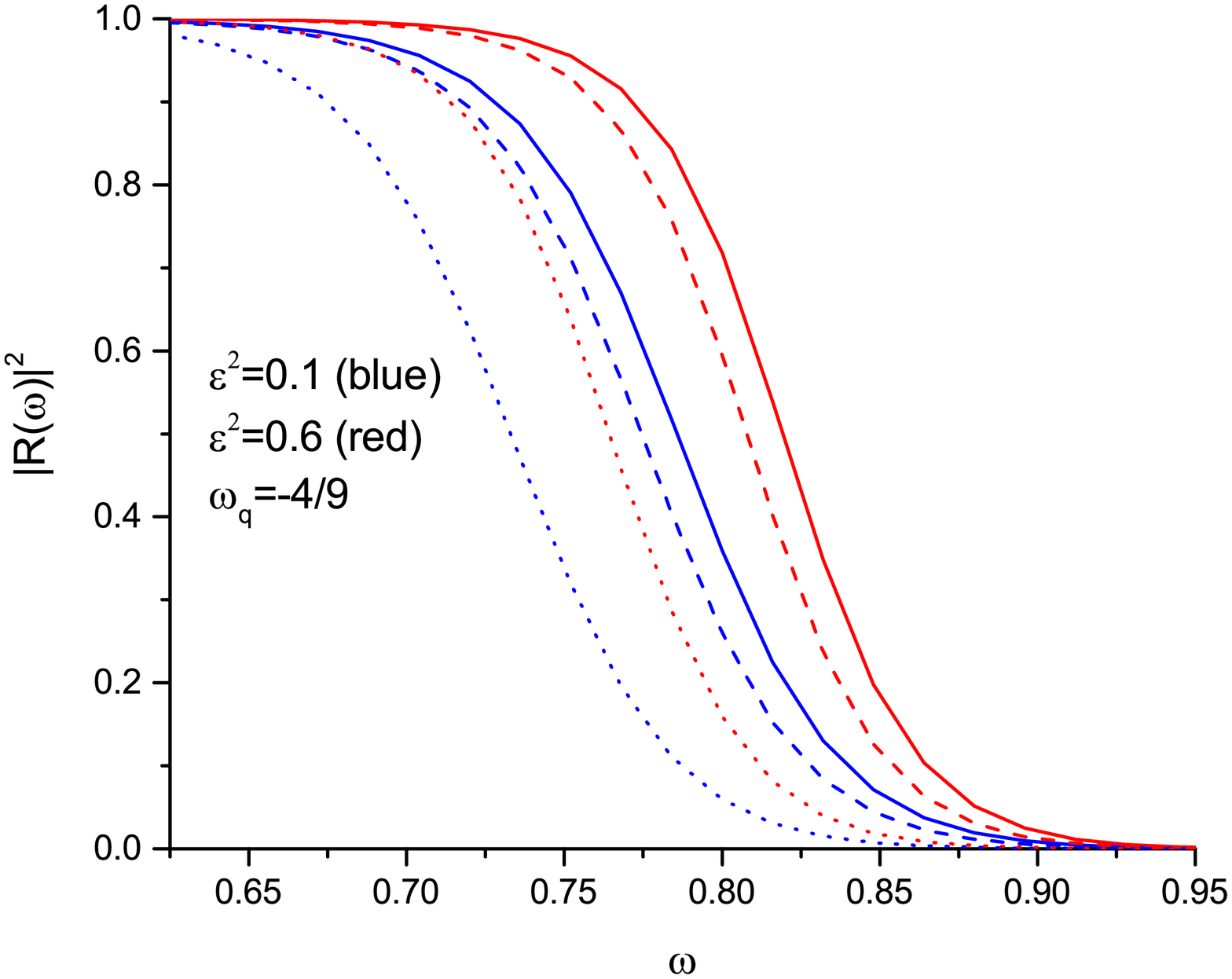}
\caption{The plot shows the reflection coefficients of the scattered scalar (solid), electromagnetic (dashed) and gravitational (dotted) wave for $l=4$, $c=0.05$ and $n=0$.}
\label{re}
\end{figure}

\begin{figure}[!h]
\centering
\includegraphics[scale=0.31]{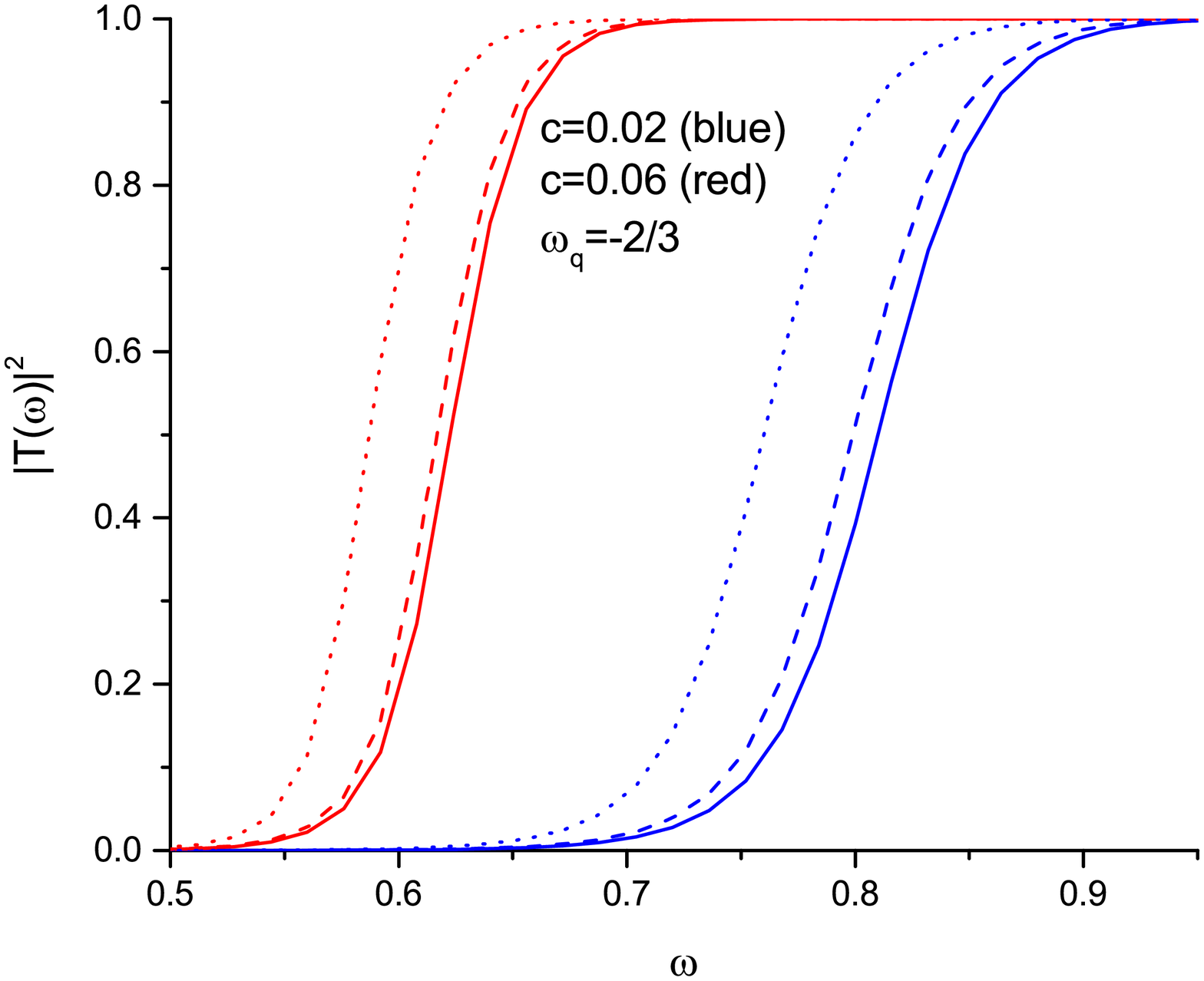}
\includegraphics[scale=0.31]{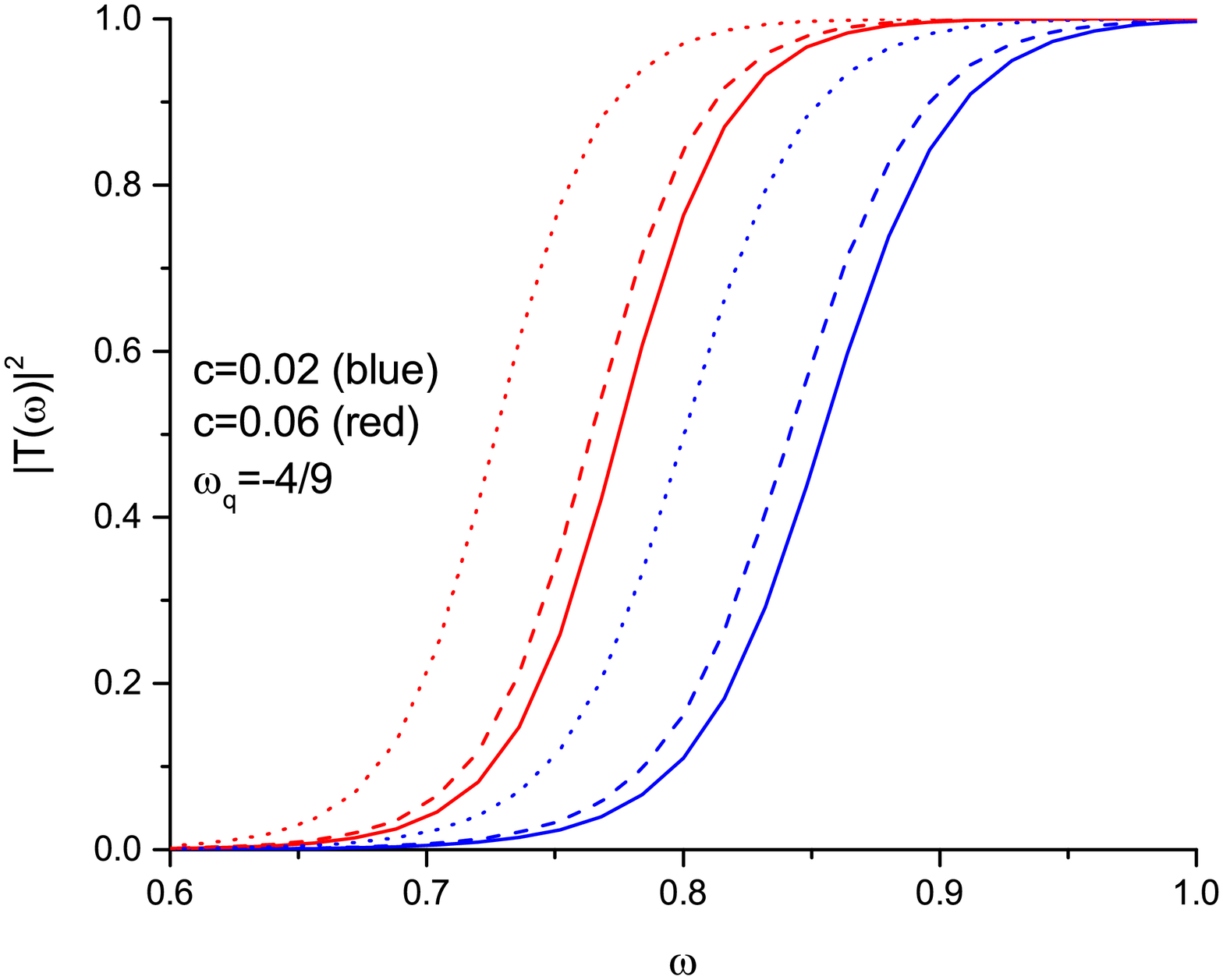}
\caption{The plot shows the transmission coefficients of the scattered scalar (solid), electromagnetic (dashed) and gravitational (dotted) wave for $l=4$, $\epsilon^2=0.2$ and $n=0$.}
\label{tc}
\end{figure}

\begin{figure}[!h]
\centering
\includegraphics[scale=0.31]{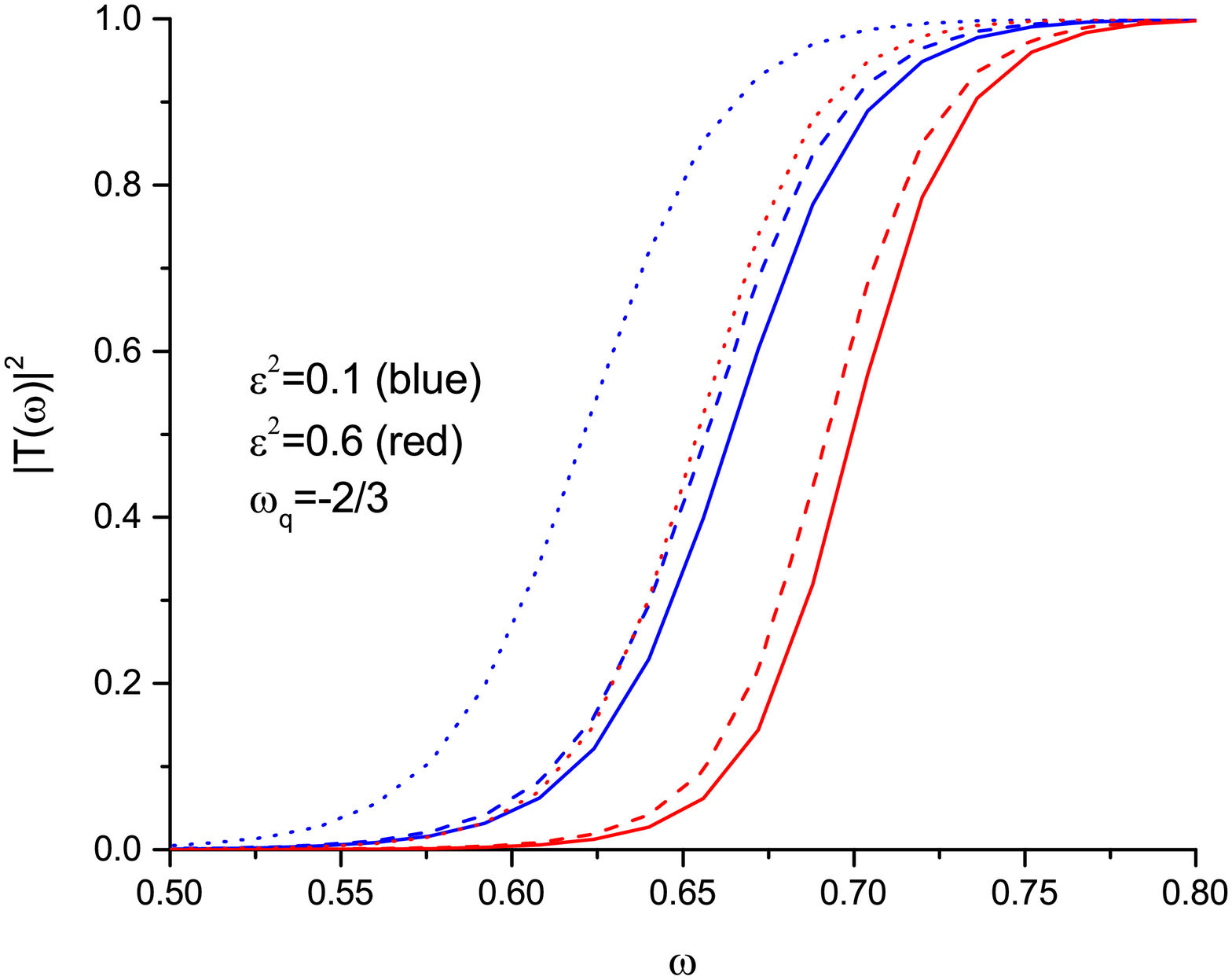}
\includegraphics[scale=0.31]{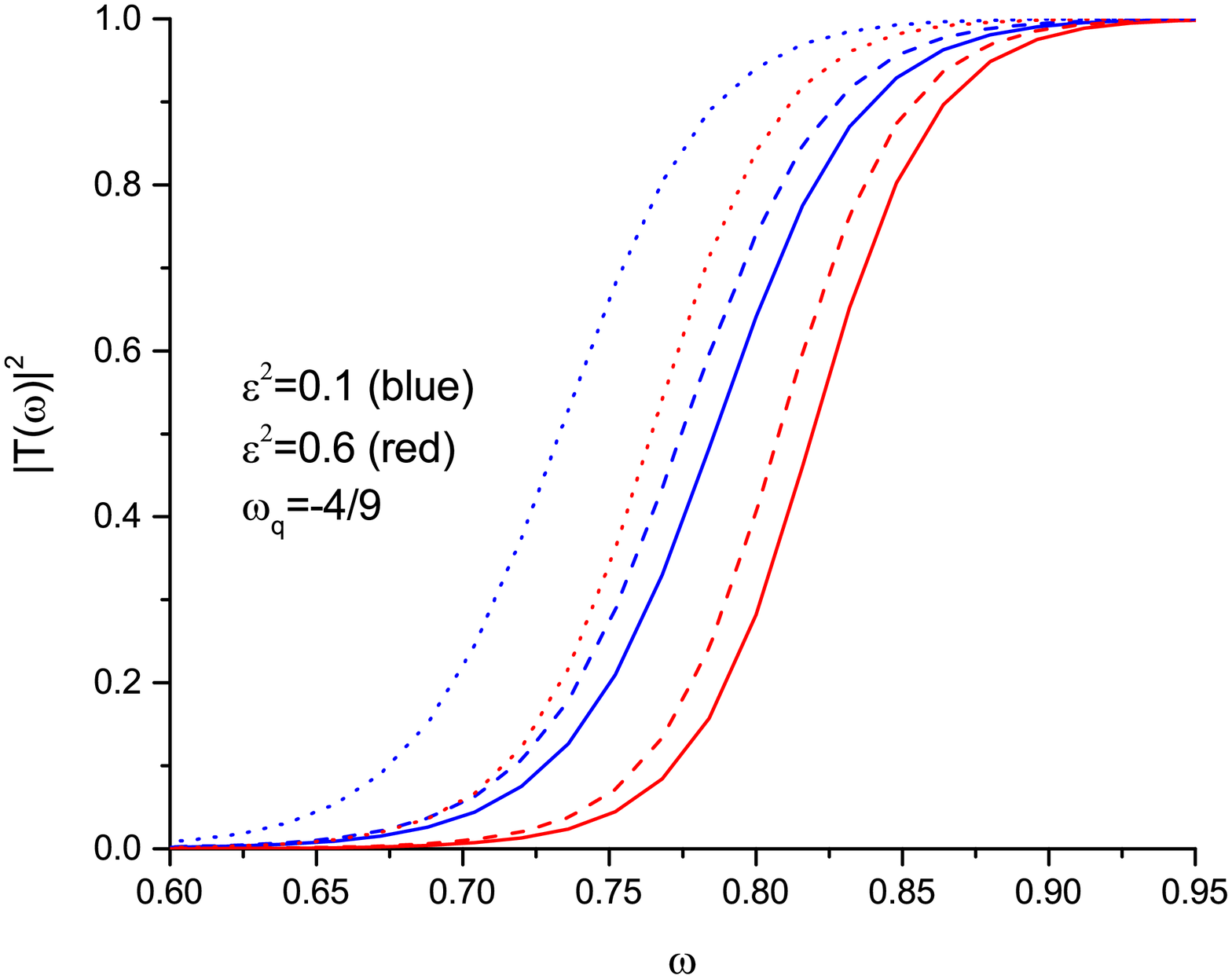}
\caption{The plot shows the transmission coefficients of the scattered scalar (solid), electromagnetic (dashed) and gravitational (dotted) wave for $l=4$, $c=0.05$ and $n=0$.}
\label{te}
\end{figure}

The numerical results of the reflection and transmission coefficients for the corresponding HBH--$\omega_q$ in terms of different values of parameters are shown in Figs. \ref{rc}--\ref{te}. Fig. \ref{rc} shows how the reflection coefficient decreases with an increasing $c$, however, in the case $\omega_q=-2/3$, the effect is more noticeable than in the case $\omega_q=-4/9$. On the other hand, the $|R(\omega)|^2$ increases as $\epsilon^2$ also increases, as is shown in Fig. \ref{re}. In general we can mention that $|R(\omega)|^2 _{sc}> |R(\omega)|^2 _{elec}>|R(\omega)|^2 _{grav}$ and that when the presence of quintessence increases, the reflection occurs in values minors de $\omega$.

Fig. \ref{tc} shows how the transmission coefficient decreases with the increase in $c$, while $|T(\omega)|^2$ increases with increasing values of $\varepsilon$ (see from Fig. \ref{te}). For scalar, electromagnetic and gravitational perturbations the behaviors are similar for $|R(\omega)|^2$ and $|T(\omega)|^2$.  
           
\section{Conclusions}

We analyzed the metric of the Hayward BH surrounded by quintessence, we considered the term $\epsilon$, the quintessence parameter ($c$), and radial distance in mass units for the analysis of horizons and the extreme case, also we presented how $\epsilon$ depends on the values of $c$.

We focus on the particular cases $\omega=-2/3$ and $\omega=-4/9$ that enable a relatively simple treatment of Hayward Black Holes properties with quintessence.  

Mainly, we studied the QNM of the scalar, electromagnetic and gravitational perturbations in the HBH--$\omega_q$ using the third--order WKB method. Results have shown that an increase in the normalization factor $c$ implies a monotonic decrease of the real and imaginary parts of the QNM frequency. On the other hand, we can see that with an increase in the parameter $\varepsilon$, the real and imaginary parts of the QNM frequencies increase. It is worth mentioning that these characteristics observed contain similar behavior for $\omega_q=-2/3$ and $\omega_q=-4/9$. The results of calculations are given in the Tables. \ref{scvsc}--\ref{gravsep}.   

We find that in the Hayward BH surrounded by quintessence, the real part of the QNM is always largest for the scalar fields and smallest for the gravitational fields. However, the roles are swapped for the damping of the QNM. In this case, the imaginary part of the QNM is all the time smallest for the scalar fields and most significant for the gravitational fields.     

From the Figs. \ref{fmivsc} and \ref{fmivse} we can figure out the stability for all perturbations considered for both values of the state parameter $\omega_q=-2/3$ and $\omega_q=-4/9$. In particular, for the case $\omega_q=-2/3$ there is no noticeable difference in the damping rate behavior as a function of the parameter $c$ for all the three types of perturbations.    

The greybody factor has been calculated by applying the third--order WKB approach for all three different types of perturbations. The greybody factor decreases with an increasing $c$ while the transmission and reflection coefficients increase with an increase in $\epsilon$, i.e., the probability of the wave transmission through the potential barrier depends inversely on the maximum of the effective potential, and this behavior can be explained from the Figs. \ref{vscw}--\ref{vslw}. Thus, an increase in the value of the $c$ weakens the potential barrier in relation, and hence the transmission coefficient increases. On the other hand, the effective potential increases with an increase in $\epsilon$, and the transmission coefficient decreases. This behavior is similar for different perturbations.

\section*{ACKNOWLEDGMENT}

The authors acknowledge the financial support from PROMEP project UAEH--CA--108 and  SNI--CONACYT, M\'exico.

\bibliographystyle{unsrt}
\bibliography{bibliografia}

\begin{thebibliography}{10}

\bibitem{LIGOScientific:2016sjg}
B.~P. Abbott et~al.
\newblock {GW151226: Observation of Gravitational Waves from a 22-Solar-Mass
  Binary Black Hole Coalescence}.
\newblock {\em Phys. Rev. Lett.}, 116(24):241103, 2016.

\bibitem{Regge:1957td}
Tullio Regge and John~A. Wheeler.
\newblock {Stability of a Schwarzschild singularity}.
\newblock {\em Phys. Rev.}, 108:1063--1069, 1957.

\bibitem{Zerilli:1970se}
Frank~J. Zerilli.
\newblock {Effective potential for even parity Regge-Wheeler gravitational
  perturbation equations}.
\newblock {\em Phys. Rev. Lett.}, 24:737--738, 1970.

\bibitem{Teukolsky:1972my}
S.~A. Teukolsky.
\newblock {Rotating black holes - separable wave equations for gravitational
  and electromagnetic perturbations}.
\newblock {\em Phys. Rev. Lett.}, 29:1114--1118, 1972.

\bibitem{Berti:2009kk}
Emanuele Berti, Vitor Cardoso, and Andrei~O. Starinets.
\newblock {Quasinormal modes of black holes and black branes}.
\newblock {\em Class. Quant. Grav.}, 26:163001, 2009.

\bibitem{Konoplya:2011qq}
R.~A. Konoplya and A.~Zhidenko.
\newblock {Quasinormal modes of black holes: From astrophysics to string
  theory}.
\newblock {\em Rev. Mod. Phys.}, 83:793--836, 2011.

\bibitem{Percival:2020skc}
Jake Percival and Sam~R. Dolan.
\newblock {Quasinormal modes of massive vector fields on the Kerr spacetime}.
\newblock {\em Phys. Rev. D}, 102(10):104055, 2020.

\bibitem{Ma:2020qkd}
Hong Ma and Jin Li.
\newblock {Dirac quasinormal modes of Born-Infeld black hole spacetimes}.
\newblock {\em Chin. Phys. C}, 44(9):095102, 2020.

\bibitem{Schutz:1985km}
Bernard~F. Schutz and Clifford~M. Will.
\newblock {BLACK HOLE NORMAL MODES: A SEMIANALYTIC APPROACH}.
\newblock {\em Astrophys. J. Lett.}, 291:L33--L36, 1985.

\bibitem{Cho:2011sf}
H.~T. Cho, A.~S. Cornell, Jason Doukas, T.~R. Huang, and Wade Naylor.
\newblock {A New Approach to Black Hole Quasinormal Modes: A Review of the
  Asymptotic Iteration Method}.
\newblock {\em Adv. Math. Phys.}, 2012:281705, 2012.

\bibitem{Cardoso:2008bp}
Vitor Cardoso, Alex~S. Miranda, Emanuele Berti, Helvi Witek, and Vilson~T.
  Zanchin.
\newblock {Geodesic stability, Lyapunov exponents and quasinormal modes}.
\newblock {\em Phys. Rev. D}, 79:064016, 2009.

\bibitem{Devi:2020uac}
Saraswati Devi, Rittick Roy, and Sayan Chakrabarti.
\newblock {Quasinormal modes and greybody factors of the novel four dimensional
  Gauss\textendash{}Bonnet black holes in asymptotically de Sitter space time:
  scalar, electromagnetic and Dirac perturbations}.
\newblock {\em Eur. Phys. J. C}, 80(8):760, 2020.

\bibitem{Toshmatov:2018ell}
Bobir Toshmatov, Zden\v{e}k Stuchl\'\i{}k, and Bobomurat Ahmedov.
\newblock {Electromagnetic perturbations of black holes in general relativity
  coupled to nonlinear electrodynamics: Polar perturbations}.
\newblock {\em Phys. Rev. D}, 98(8):085021, 2018.

\bibitem{Toshmatov:2018tyo}
Bobir Toshmatov, Zden\v{e}k Stuchl\'\i{}k, Jan Schee, and Bobomurat Ahmedov.
\newblock {Electromagnetic perturbations of black holes in general relativity
  coupled to nonlinear electrodynamics}.
\newblock {\em Phys. Rev. D}, 97(8):084058, 2018.

\bibitem{Breton:2016mqh}
N.~Breton and L.~A. Lopez.
\newblock {Quasinormal modes of nonlinear electromagnetic black holes from
  unstable null geodesics}.
\newblock {\em Phys. Rev. D}, 94(10):104008, 2016.

\bibitem{Toshmatov:2015wga}
Bobir Toshmatov, Ahmadjon Abdujabbarov, Zden\v{e}k Stuchl\'\i{}k, and Bobomurat
  Ahmedov.
\newblock {Quasinormal modes of test fields around regular black holes}.
\newblock {\em Phys. Rev. D}, 91(8):083008, 2015.

\bibitem{Capozziello_2006}
S~Capozziello, V~F Cardone, E~Piedipalumbo, and C~Rubano.
\newblock Dark energy exponential potential models as curvature quintessence.
\newblock {\em Classical and Quantum Gravity}, 23(4):1205--1216, feb 2006.

\bibitem{PhysRevLett.81.3067}
Sean~M. Carroll.
\newblock Quintessence and the rest of the world: Suppressing long-range
  interactions.
\newblock {\em Phys. Rev. Lett.}, 81:3067--3070, Oct 1998.

\bibitem{Kiselev:2002dx}
V.~V. Kiselev.
\newblock {Quintessence and black holes}.
\newblock {\em Class. Quant. Grav.}, 20:1187--1198, 2003.

\bibitem{Ghaderi:2017wvl}
K.~Ghaderi.
\newblock {Geodesics of black holes with dark energy}.
\newblock {\em Astrophys. Space Sci.}, 362(12):218, 2017.

\bibitem{Fernando:2012ue}
Sharmanthie Fernando.
\newblock {Schwarzschild black hole surrounded by quintessence: Null
  geodesics}.
\newblock {\em Gen. Rel. Grav.}, 44:1857--1879, 2012.

\bibitem{Malakolkalami:2015tsa}
B.~Malakolkalami and K.~Ghaderi.
\newblock {The null geodesics of the Reissner--Nordström black hole surrounded
  by quintessence}.
\newblock {\em Mod. Phys. Lett. A}, 30(10):1550049, 2015.

\bibitem{Ghosh:2016ddh}
Sushant~G. Ghosh, Muhammed Amir, and Sunil~D. Maharaj.
\newblock {Quintessence background for 5D
  Einstein\textendash{}Gauss\textendash{}Bonnet black holes}.
\newblock {\em Eur. Phys. J. C}, 77(8):530, 2017.

\bibitem{Saleh:2018hba}
Mahamat Saleh, Bouetou~Bouetou Thomas, and Timoleon~Crepin Kofane.
\newblock {Quasinormal modes of gravitational perturbation around regular
  Bardeen black hole surrounded by quintessence}.
\newblock {\em Eur. Phys. J. C}, 78(4):325, 2018.

\bibitem{Polchinski:1989ae}
Joseph Polchinski.
\newblock {Decoupling Versus Excluded Volume or Return of the Giant Wormholes}.
\newblock {\em Nucl. Phys.}, B325:619--630, 1989.

\bibitem{Hayward:2005gi}
Sean~A. Hayward.
\newblock {Formation and evaporation of regular black holes}.
\newblock {\em Phys. Rev. Lett.}, 96:031103, 2006.

\bibitem{Amir:2015pja}
Muhammed Amir and Sushant~G. Ghosh.
\newblock {Rotating Hayward's regular black hole as particle accelerator}.
\newblock {\em JHEP}, 07:015, 2015.

\bibitem{Frolov:2016pav}
Valeri~P. Frolov.
\newblock {Notes on nonsingular models of black holes}.
\newblock {\em Phys. Rev.}, D94(10):104056, 2016.

\bibitem{Lopez}
L.A. López and V.~Hinojosa.
\newblock Quasinormal modes of charged regular black hole.
\newblock {\em Canadian Journal of Physics}, 99(1):44--48, 2021.

\bibitem{Lin:2013ofa}
Kai Lin, Jin Li, and Shuzheng Yang.
\newblock {Quasinormal Modes of Hayward Regular Black Hole}.
\newblock {\em Int. J. Theor. Phys.}, 52:3771--3778, 2013.

\bibitem{Pedraza:2020uuy}
Omar Pedraza, L.~A. L\'opez, R.~Arceo, and I.~Cabrera-Munguia.
\newblock {Geodesics of Hayward black hole surrounded by quintessence}.
\newblock {\em Gen. Rel. Grav.}, 53(3):24, 2021.

\bibitem{Rizwan:2018lht}
Muhammad Rizwan, Mubasher Jamil, and Anzhong Wang.
\newblock {Distinguishing a rotating Kiselev black hole from a naked
  singularity using the spin precession of a test gyroscope}.
\newblock {\em Phys. Rev. D}, 98(2):024015, 2018.
\newblock [Erratum: Phys.Rev.D 100, 029902 (2019)].

\bibitem{Medved:2004tp}
A.~J.~M. Medved, Damien Martin, and Matt Visser.
\newblock {Dirty black holes: Symmetries at stationary nonstatic horizons}.
\newblock {\em Phys. Rev. D}, 70:024009, 2004.

\bibitem{PhysRevD.71.124033}
Hidefumi Nomura and Takashi Tamaki.
\newblock Continuous area spectrum of a regular black hole.
\newblock {\em Phys. Rev. D}, 71:124033, Jun 2005.

\bibitem{Iyer:1986np}
Sai Iyer and Clifford~M. Will.
\newblock {Black Hole Normal Modes: A {WKB} Approach. 1. Foundations and
  Application of a Higher Order {WKB} Analysis of Potential Barrier
  Scattering}.
\newblock {\em Phys. Rev. D}, 35:3621, 1987.

\bibitem{Konoplya:2003ii}
R.~A. Konoplya.
\newblock {Quasinormal behavior of the d-dimensional Schwarzschild black hole
  and higher order WKB approach}.
\newblock {\em Phys. Rev. D}, 68:024018, 2003.

\bibitem{Konoplya:2009hv}
R.~A. Konoplya and A.~Zhidenko.
\newblock {Holographic conductivity of zero temperature superconductors}.
\newblock {\em Phys. Lett. B}, 686:199--206, 2010.

\end{thebibliography}

\end{document}